\documentclass[prd,nofootinbib,superscriptaddress,notitlepage,10pt,aps]{revtex4-1}
\usepackage{epsfig,amsmath,amssymb,slashed}
\usepackage[utf8]{inputenc}
\usepackage{hyperref}
\usepackage{amsthm}
\usepackage{xcolor}
\usepackage{bm}
\usepackage{combelow}
\usepackage[normalem]{ulem}

\renewcommand{\tilde}{\widetilde}

\newcommand{\ket}[1]{|#1\rangle}

\newcommand{\wad}[2]{\widehat a^\dagger_{#2}(#1)}
\newcommand{\wa}[2]{\widehat a_{#2}(#1)}

\newcommand{\di}{{\rm d}}
\newcommand{\Tr}{{\rm Tr}}
\newcommand{\ii}{i}

\newcommand{\abs}[1]{\left| #1\right|}

\def\wQ{{\widehat Q}}
\def\wP{{\widehat P}}
\def\wJ{{\widehat J}}
\def\wS{{\widehat S}}

\def\wh{{\widehat h}}

\def\wPi{{\widehat{\Pi}}}

\def\wrho{{\widehat{\rho}}}

\newcommand{\tr}{{\rm tr}}  
  
\newcommand{\e}{{\rm e}}

\newcommand{\be}{\begin{equation}}
\newcommand{\ee}{\end{equation}}                                                                               
\newcommand{\bea}{\begin{eqnarray}}
\newcommand{\eea}{\end{eqnarray}}

\begin{document}

\title{Exact spin polarization of massive and massless particles in relativistic fluids 
at global equilibrium}

\author{Andrea Palermo}\affiliation{Universit\`a di 
 Firenze and INFN Sezione di Firenze, Via G. Sansone 1, 
	I-50019 Sesto Fiorentino (Florence), Italy}\affiliation{
Institut f\"ur Theoretische Physik, Johann Wolfgang Goethe-Universit\"at, Max-von-Laue-Straße 1,
D-60438 Frankfurt am Main, Germany}
\author{Francesco Becattini}\affiliation{Universit\`a di 
 Firenze and INFN Sezione di Firenze, Via G. Sansone 1, 
	I-50019 Sesto Fiorentino (Florence), Italy}

\begin{abstract}
We present the exact form of the spin polarization vector and the spin density matrix of massive
and massless free particles of any spin and helicity at general global equilibrium in a relativistic
fluid with non-vanishing thermal vorticity, thus extending the known expression at the linear order. 
The exact form is obtained by means of the analytic continuation of the relativistic density operator 
to imaginary thermal vorticity and the resummation of the obtained series. The phenomenological implications 
for the polarization of the $\Lambda$ hyperon in relativistic heavy-ion collisions are addressed.
\end{abstract}

\maketitle

\section{Introduction}

Following the evidence of spin polarization of the $\Lambda$ hyperon \cite{STAR:2017ckg}, spin physics in 
relativistic heavy ion collisions has become a very active research field both at the experimental  \cite{STAR:2019erd,STAR:2020xbm,STAR:2023eck,ALICE:2021pzu,ALICE:2019onw,Kornas:2020qzi} and theoretical level \cite{Florkowski:2017ruc,Florkowski:2018fap,Weickgenannt:2019dks,Weickgenannt:2020aaf,Huang:2020xyr,Gao:2020lxh,Becattini:2020ngo,Gao:2020vbh,Hongo:2021ona}
(see \cite{Becattini:2022zvf} for a recent review).

At local thermodynamic equilibrium in a relativistic fluid, spin polarization turns out to be a function 
of the gradients of the thermo-hydrodynamic fields, particularly the gradient of the four-temperature 
vector $\beta$ which is related to proper temperature and four-velocity of the fluid by $\beta^\mu=(1/T)u^\mu$. 
The gradients of $\beta$ include an anti-symmetric part called thermal vorticity $\varpi$:
\begin{equation}\label{thvort}
    \varpi_{\mu\nu}=-\frac{1}{2}(\partial_\mu\beta_\nu-\partial_\nu\beta_\mu),
\end{equation}
and a symmetric part called \emph{thermal shear}. It has been recently found out that the thermal shear induces
a significant polarization in relativistic nuclear collisions \cite{Liu:2021uhn,Fu:2021pok,
Becattini:2021suc,Becattini:2021iol,Yi:2021ryh}. At global equilibrium, however, thermal shear vanishes because the field $\beta$ must
become a Killing vector \cite{DeGroot:1980dk,Becattini:2012tc}, and only a constant thermal vorticity survives.  

For a spin-$1/2$ free Dirac field, the expression of the spin polarization vector at the leading order
in thermal vorticity was derived in \cite{Becattini:2013fla}:
\begin{equation}\label{eq:linear spin vector vort}
    S^\mu(p)=-\frac{1}{8m}\epsilon^{\mu\nu\rho\sigma}p_\sigma\frac{\int\di\Sigma\cdot p 
    \,\varpi_{\nu\rho}n_F(1-n_F)}{\int\di\Sigma\cdot p \,n_F},
\end{equation}
where 
\begin{equation}\label{eq: fermi-dirac}
n_F=\frac{1}{\exp(\beta\cdot p -\zeta)+1},
\end{equation} 
is the Fermi-Dirac distribution function and $\zeta=\mu/T$ is the ratio between the chemical potential 
and the temperature. The integrals in \eqref{eq:linear spin vector vort} must be performed over the 
freeze-out hypersurface in relativistic heavy ion physics. The equation \eqref{eq:linear spin vector vort} 
was confirmed in other derivations \cite{Fang:2016vpj,Shi:2020htn,Becattini:2021suc}  and it is a good approximation 
for small values of $\varpi$. Yet, very little is known about higher-order terms, not even at global 
equilibrium where thermal vorticity is a constant.

In this work, we obtain the exact expression of the spin density matrix and spin polarization vector 
of massive and massless free fields of any spin at general global equilibrium with non-vanishing thermal 
vorticity, by means of the analytic continuation of the density operator proposed in 
refs.~\cite{Becattini:2020qol, Palermo:2021hlf}. We will first derive those expressions for free 
Dirac fermions by using the covariant Wigner function formalism, thereafter showing that they are a 
special case of a more general formula applying to any spin. 
Finally, we compare the newly found expressions to the linear approximation \eqref{eq:linear spin vector vort} 
which is commonly used in phenomenological studies, in order to assess the impact of higher-order 
corrections in thermal vorticity for the spin polarization measurements in relativistic heavy ion 
physics.

\subsection*{Notations}

We use the natural units, with $\hbar=c=K=1$. The Minkowskian metric tensor $g$ is 
${\rm diag}(1,-1,-1,-1)$ and repeated indices are assumed to be saturated; for the Levi-Civita symbol, 
we use the convention $\epsilon^{0123}=1$. Three vectors are denoted with bold symbols, for example 
$\bm{v}$. This notation corresponds to the contravariant space components of the corresponding four-vector, 
such that $v^\mu = (v^0,\bm{v})$. Unit vectors are denoted with a small upper hat, e.g. $\hat p$.
The notation $ a \cdot b$ is sometimes used for the scalar product of four-vectors and
$X : Y$ for the double contraction of tensors, i.e. $X : Y = X^{\mu\nu} Y_{\mu\nu}$.
\\
Operators in Hilbert space will be denoted by a wide upper hat, e.g. $\widehat H$, except the Dirac 
field operator which is denoted by a $\Psi$. The symbol $\Tr$ will stand for the trace over the Hilbert 
space of quantum states, while $\tr$ is the trace over a finite-dimensional vector space. We will use the notation 
$\langle\bullet\rangle=\Tr(\wrho \;\bullet)$ for thermal expectation values, $\wrho$ being the density 
operator.

\section{Spin and helicity in quantum field theory}
\label{sec: Polarization QFT}

In a quantum relativistic framework, the spin polarization vector is defined as the expectation value
of the Pauli-Lubanski (PL) operator \cite{tung1985group}:
\begin{equation}\label{eq: def PL vector}
    \wPi^\mu=-\frac{1}{2}\epsilon^{\mu\nu\rho\sigma}\wJ_{\nu\rho}\wP_\sigma,
\end{equation}
where $\wJ^{\mu\nu}$ and $\wP^\mu$ are the angular momentum-boost and the four-momentum operators 
respectively. From the Lie algebra of the Poincaré group, it follows that the PL operator fulfills 
these relations:
\begin{subequations}\label{eq:algebra W}
\begin{align}
&[\wPi^\mu,\wP^\nu]=0,\label{eq: w commutes with p}\\    
&[\wPi^\mu,\wPi^\nu]=-\ii \epsilon^{\mu\nu\rho\sigma}\wPi_\rho\wP_\sigma,\label{eq: commutations w}\\
&\wPi\cdot\wP=0. \label{eq s ortogonale a p}
\end{align}
\end{subequations}
The restriction of the PL operator to the one-particle states with definite momentum $\ket{p}$ is defined 
as $\wPi(p)$:
\begin{align}
\wPi^\mu(p)=-\frac{1}{2}\epsilon^{\mu\nu\rho\sigma}\wJ_{\nu\rho} p_\sigma.
\end{align}
This operator generates the so-called little group of $p$, that is the group of Lorentz transformation 
leaving $p^\mu$ invariant, and plays a crucial role in the definition of the spin and helicity operators. 
We first notice that, due to eq. \eqref{eq s ortogonale a p}, 
$\wPi(p)$ can be decomposed along the directions perpendicular to $p^\mu$. Such decomposition is 
different for massive and massless states, because in the latter case $p^\mu$ is orthogonal to itself. 

In the massive case one can define a $p$-dependent spin operator:
\begin{equation}\label{eq: def spin operator}
    \wS^{\mu}(p)=\frac{\wPi^\mu(p)}{m}.
\end{equation}
Due to equation \eqref{eq s ortogonale a p}, and since $p^\mu$ is time-like, the decomposition of the 
spin operator can be made along three orthogonal, normalized space-like vectors $n^{\mu}_i(p)$, such that 
$n_i(p)\cdot p=0$ and $n_i\cdot n_j=-\delta_{ij}$. These vectors, along with $p$, make a momentum-dependent 
orthogonal basis of the Minkowski space-time. For a particle at rest, we have $p=(m,0,0,0)$ and we define the $n_i$
to coincide with the conventional basis vector $\e_i$. In this special frame, vectors will be denoted by 
Gothic letters, i.e. $\mathfrak{p}^\mu=(m,0,0,0)$ - which will be henceforth referred to as {\em standard
vector} - and $\mathfrak{n}_i = \e_i$. Furthermore, a so-called {\em standard Lorentz transformation} $[p]$ is
introduced, which transforms the conventional basis to the particle basis. Explicitly:
\begin{align*}
    p^\mu &=[p]^{\mu}_{\ \nu}\mathfrak{p}^\nu,
    &&n_i^\mu(p)=[p]^{\mu}_{\ \nu}\mathfrak{n}^\nu_i.
\end{align*}

The decomposition of the spin operator along the tetrad $\{p,n_1,n_2,n_3\}$ reads:
\begin{align}\label{eq:decomposition S massive}
    \wS^{\mu}(p)=\sum_{i=1}^3 \wS_i(p)n_i^\mu(p) \implies \wS_i(p)=-\wS(p)\cdot n_i(p).
\end{align}
It is well known that the components $\wS_i(p)$ fulfill a SO(3) Lie algebra and are related to the generators 
of rotations. Particularly:
\begin{equation}\label{eq: PL is J}
\wS_i(\mathfrak{p})= \widehat{{\rm J}}^i,    
\end{equation}
$\widehat{{\rm J}}^i$ being the $i$-th (contravariant) component of the  angular momentum operator.

In a statistical system, the mean spin polarization vector of a massive particle with momentum $p$ 
can be obtained with the formula \cite{becattini2020polarization}:
\begin{align}\label{eq:spin vect tr J}
S^\mu(p)=\sum_{i=1}^3 [p]^\mu_{\ i}\tr\left(\Theta(p) D^S(\mathrm{J}^i)\right),
\end{align}
where $D^S(\mathrm{J}^i)$ is the i-th generator of the rotation group in the 
spin $S$ representation and $\tr$ denotes the trace on the $(2S+1)$-dimensional spin space.
The matrix $\Theta(p)$ is the spin density matrix and in a quantum field
theoretical framework reads:
\begin{align}\label{definition theta field theory}
    \Theta_{sr}(p)=\frac{\Tr\left(\wrho\,\wad{p}{r}\wa{p}{s}\right)}
    {\sum_{l}\Tr\left(\wrho\,\wad{p}{l}\wa{p}{l}\right)}=
    \frac{\langle\wad{p}{r}\wa{p}{s}\rangle}{\sum_l\langle\wad{p}{l}\wa{p}{l}\rangle},
\end{align}
where $\wad{p}{s}$ and $\wa{p}{s}$ are the creation and annihilation operators for particles with 
momentum $p$ and spin (helicity) $s$. 

For light-like states, the decomposition of the PL operator is different and the formula 
\eqref{eq:spin vect tr J} no longer holds. The four-momentum $p$ can be included in a non-orthonormal 
basis of Minkowski space-time defined by the tetrad $\{p,q,n_1,n_2\}$, where $q$ is a light-like vector 
\emph{not} orthogonal to $p$ and $n_{1,2}$ are two normalized space-like vectors orthogonal to $p$, 
$q$ and to each other (i.e., $n_i(p)\cdot p = n_i(p)\cdot q =0$, $n_i(p)\cdot n_j(p)=-\delta_{ij}$). 
Similarly to the massive case, these vectors can be written as Lorentz-transformed of a basis of
standard vectors. They are, for some $\kappa>0$:
$$
\mathfrak{p}^\mu=(\kappa,0,0,\kappa), \qquad 
\mathfrak{q}^\mu=(\kappa,0,0,-\kappa), \qquad \mathfrak{n}_{i}^{\mu}=\delta^\mu_i.
$$
Likewise, a standard Lorentz transformation $[p]$ is introduced (a typical choice being a boost 
along the $z$ direction followed by a rotation around the $\hat {\bf k} \times \hat{\bf{p}}$ axis),  
turning the standard basis into the particle basis:
$$
p^\mu=[p]^{\mu}_{\ \nu}\mathfrak{p}^\nu,\qquad\qquad
q^\mu=[p]^{\mu}_{\ \nu}\mathfrak{q}^\nu,\qquad\qquad
n^\mu_i(p)=[p]^{\mu}_{\ \nu}\mathfrak{n}_i^\nu.
$$
Taking into account $\wPi(p)\cdot p=0$, the component of the PL vector operator restricted
to single particle states along $q^\mu$ vanishes and one has:
\begin{align}\label{eq: decomposition of S}
    \wPi^\mu(p)=\wh(p)p^\mu+\wPi_1(p)n^\mu_1(p)+\wPi_2(p)n^\mu_2(p).
\end{align}
Using equations \eqref{eq:algebra W}, it is possible to show that the components of the above 
decomposition obey the following commutation rules:
\begin{subequations}\label{eq:algebra h massless}
\begin{align}
    &[\wh(p),\wPi_1(p)]=i\wPi_2(p),\\
    &[\wh(p),\wPi_2(p)]=-i\wPi_1(p),\\
    &[\wPi_1(p),\wPi_2(p)]=0.
\end{align}
\end{subequations}
This is the algebra of the euclidean group in two dimensions, ISO(2). The algebra is semi-simple, 
including an abelian sub-algebra generated by $\wPi_{1,2}(p)$. It is well known that actual 
physical states are such that:
\begin{align}\label{eq s1|p>=0}
    \wPi_1(p)|p\rangle=\wPi_2(p)|p\rangle=0.
\end{align}
A straightforward consequence of \eqref{eq s1|p>=0} is that the only physically relevant component 
of the PL vector is $\wh(p)$ and a basis of the Hilbert space can be chosen using the common eigenvectors 
of $\wh(p)$ and $\wP^\mu$:
\begin{subequations}\label{eq: basis construction}
    \begin{align}
    \wP^\mu|p,h\rangle&=p^\mu|p,h\rangle, \\
    \wh(p)|p,h\rangle&=h|p,h\rangle.\label{eq:helicity eigenvalue}
    \end{align}
\end{subequations}
The eigenvalue of the $\wh(p)$ operator is referred to as the \emph{helicity} of the particle. Due to 
the topology of the Lorentz group, helicity can only be integer or half-integer, and it is also known 
that helicity-S massless states only exhibit the two extremal helicity states, $S$ and $-S$ \cite{Weinberg1995qft1}. 
Contracting the eq. \eqref{eq: decomposition of S} with $q$, it can be realized that the helicity 
operator can be written as:
\begin{equation*}
     \wh(p) = \frac{\wPi(p)\cdot q}{p\cdot q}=
     -\frac{1}{2p\cdot q}\epsilon^{\mu\nu\rho\sigma}
     \wJ_{\nu\rho}p_\sigma q_\mu.
\end{equation*}
For a single relativistic particle we thus have, if $\wrho$ is the single-particle density operator:
\begin{align}\label{polvectmass0}
    \Pi^\mu(p)= \sum_h\langle p, h|\wPi^\mu(p) \wrho |p,h\rangle
    = p^\mu\sum_h h\langle p,h|\wrho |p,h\rangle = p^\mu \sum_h h \Theta(p)_{hh} , 
\end{align}
where we used the decomposition \eqref{eq: decomposition of S} and the equation \eqref{eq s1|p>=0}.
The sum in equation \eqref{polvectmass0}, as it is known, has just two terms, i.e. $h = -S$ and $h=S$.
In a statistical system, the mean polarization vector of a particle with momentum $p$ is obtained 
from the equation \eqref{eq: decomposition of S} with the spin density matrix:
\begin{equation}\label{s in terms of aa}
    \Theta(p)_{hh} =\frac{\langle\wad{p}{h}\wa{p}{h}\rangle}
    {\sum_{l=\pm S} \langle\wad{p}{l}\wa{p}{l}\rangle},
\end{equation}
which can be interpreted as the fraction of particles with helicity $h$. Altogether, the
mean PL vector reads:
\begin{align}\label{polvectmass1}
    \Pi^\mu(p) = p^\mu \frac{\sum_{h=\pm S} h \langle\wad{p}{h}\wa{p}{h}\rangle}
    {\sum_{l=\pm S} \langle\wad{p}{l}\wa{p}{l}\rangle}.
\end{align}
%
\section{Spin polarization of Dirac fermions and the Wigner function}\label{sec: spin fermions wigner}

The Wigner function is a useful tool in spin polarization studies. For a non-interacting Dirac field, 
either massive or massless, the Wigner function is defined as \cite{DeGroot:1980dk}:
\begin{equation*}
    W(x,k)=-\frac{1}{(2\pi)^4}\int\di^4y \, \e^{-ik\cdot y}\,
    \Tr \left( \wrho :\Psi\left({\small x-\frac{y}{2}}\right)\overline{\Psi} \left({\small x+\frac{y}{2}}\right):\right), 
\end{equation*}
where the colons imply normal ordering, $\Psi$ is the Dirac field operator (the Dirac adjoint being 
$\overline{\Psi}=\Psi^\dagger \gamma^0$) and $\wrho$ is the density operator. Notice that the pseudo-momentum $k$ 
is not in general on-shell. In fact, depending on $k$, it is possible to decompose the Wigner function into
particle, antiparticle, and spacelike components, denoted as $W_+$, $W_-$ and $W_S$ respectively:
\begin{equation*}
    W(x,k)=W_+(x,k)\theta(k^2)\theta(k^0)+W_-(x,k)\theta(k^2)\theta(-k^0)+W_s(x,k)\theta(-k^2).
\end{equation*}
Since the definition of the Wigner function, as well as the plane wave expansion of the non-interacting Dirac 
field,  are the same for massive and massless particles, some properties of the Wigner function are common to 
the two cases. For instance, it can be shown that:
\begin{equation}\label{property conservation}
    k^\mu \partial_\mu W_{\pm s}(x,k)=0.
\end{equation}
The above equation implies that, provided that some boundary conditions are fulfilled, $k^\mu W_{\pm s}(x,k)$ can be 
integrated over any hypersurface, and the result is independent thereof. This property makes it possible
to define an on-shell Wigner function, denoted as $w(p)$  \cite{becattini2020polarization}. Focusing on 
the particle component, one has:
\begin{equation}\label{eq: on shell wigner function}
    \frac{1}{2\varepsilon}\delta(k^0-\varepsilon)w_+(k)=\int\di\Sigma_\mu k^\mu W_+(x,k), 
\end{equation}
and, after using the plane wave expansion of the Dirac field, explicit integration leads to:
\be\label{value on shell Wigner function}
    w_+(p)=\frac{1}{2}\sum_{h,l}\langle \wad{p}{l}\wa{p}{h}\rangle u_h(p)\bar{u}_l(p).
\ee

A derivation of the formula connecting the mean spin polarization vector to the Wigner function
in the massive case was presented in ref.~\cite{becattini2020polarization}. Using the normalization of the 
spinors, $\bar{u}_s(p)u_r(p)=2m\delta_{rs}$, one gets:
\begin{equation}\label{eq: S in terms of W massive}
S^\mu(p) =\frac{1}{2}\frac{\int \di\Sigma\cdot p \; \tr(\gamma^\mu \gamma_5 W_+(x,p))}
    {\int \di\Sigma\cdot p \; \tr(W_+(x,p))},
\end{equation}
where in heavy-ions applications the integral is computed over the freeze-out hypersurface.
In fact, the derivation of a formula such as \eqref{eq: S in terms of W massive} does not trivially 
extend to the massless case, the reason being that $\bar u_s (p) u_r(p) = 0$. In this case, using the relation $\bar{u}_s(p)\gamma^\mu u_r(p)=2p^\mu\delta_{rs}$, it is easy to see:
\begin{equation}\label{eq:denominator with wigner}
    \tr(w_+(p)\gamma^\mu)=\frac{1}{2}\sum_h\langle \wad{p}{l}\wa{p}{h}\rangle \tr(u_h(p)\bar{u}_l(p)\gamma^\mu)
    =\frac{1}{2}\sum_{hl}\langle \wad{p}{l}\wa{p}{h}\rangle \bar{u}_l(p)\gamma^\mu u_h(p)= 
    p^\mu \sum_{h}\langle\wad{p}{h}\wa{p}{h}\rangle, 
\end{equation}
where the cyclicity of the trace has been used. Notice how this trace is just the denominator of the 
formula \eqref{s in terms of aa}  multiplied by $p^\mu$. Therefore, the denominator of equation 
\eqref{s in terms of aa} can be obtained by contracting the \eqref{eq:denominator with wigner} with
\emph{any} vector $v$, provided that it is not orthogonal to $p$. However, any four-vector $v$ 
can be decomposed along the basis $\{p,q,n_1,n_2\}$, as we have seen in section \ref{sec: Polarization QFT}:
$$
v^\mu=v_p p^\mu + v_q q^\mu + v_i n_i^\mu \; .
$$ 
Equation \eqref{eq:denominator with wigner} implies that only the component of $v^\mu$ along $q^\mu$ 
is relevant to invert eq. \eqref{eq:denominator with wigner}, as it is the only one contributing to 
the product $v\cdot p$. Therefore, we can choose conveniently $v=q$ to find:
\begin{equation}\label{qcontract}
    \sum_h\langle\wad{p}{h}\wa{p}{h}\rangle=\frac{\tr(w_+(k)\slashed{q})}{p\cdot q}.
\end{equation}
With the same steps of the derivation of the eq.~\eqref{eq:denominator with wigner}, it can be shown that:
\begin{equation*}
    \tr(w_+(p)\gamma^\mu\gamma_5)=2p^\mu \sum_h h \langle\wad{p}{h}\wa{p}{h}\rangle,
\end{equation*}
where we used the equation $\gamma_5 u_h(p)=2h u_h(p)$, notably applying to massless fermions. 
By contracting with $q$ the above equation and using the \eqref{qcontract}, the equation \eqref{polvectmass1} 
can be rewritten as:
\begin{equation*}
    \Pi^\mu(p)=\frac{p^\mu}{2}\frac{\tr(\slashed{q}\gamma_5 w_+(p))}{\tr(w_+(p)\slashed{q})},
\end{equation*}
and, by using the equation \eqref{eq: on shell wigner function}:
\begin{equation}\label{S as function of W}
    \Pi^\mu(p)=\frac{p^\mu}{2}\frac{\int\di\Sigma\cdot p \;\tr(\slashed{q}\gamma^5 
    W_+(x,p))}{\int \di\Sigma\cdot p \;\tr(W_+(x,p)\slashed{q})}.
\end{equation}

This formula is the corresponding of \eqref{eq: S in terms of W massive} for massless particles 
and it is the final result of this section. It should be emphasized that, in spite of its 
appearance, the equation \eqref{S as function of W} does not depend on the particular vector
$q$ chosen; this dependence cancels out, though not manifestly, if the same $q$ is used in the 
numerator and the denominator. We point out that the result \eqref{S as function of W} differs 
from others in literature \cite{Liu:2020flb} in that the polarization vector is manifestly
parallel to the four-momentum. 

\section{Exact spin polarization of Dirac fermions}\label{sec: exact pol fermions}

In a previous paper of ours \cite{Palermo:2021hlf} we derived the exact form of the Wigner function 
of free Dirac fermions in global equilibrium with non-vanishing thermal vorticity. We are now in a 
position to use that result to calculate the exact expression of the spin polarization vector of Dirac 
fermions under those conditions.

To begin with, we make a brief recap of the key concepts of global thermodynamic equilibrium in 
quantum relativistic statistical mechanics. The density operator corresponding to the most general 
global equilibrium allowed by special relativity reads:
\begin{equation}\label{eq:density op glob eq}
    \wrho=\frac{1}{Z}\exp\left[-b\cdot \wP+\frac{\varpi:\wJ}{2}+\zeta \wQ\right],
\end{equation}
where the operators $\wP$, $\wJ$ and $\wQ$ are the four-momentum, the angular momentum-boost, and the 
charge operators. The Lagrange multipliers $b^\mu$ and $\varpi^{\mu\nu}$ are a constant vector and 
a constant anti-symmetric tensor respectively. Together, they 
define the four-temperature $\beta^\mu$ as the Killing vector:
\begin{equation}\label{betakill}
    \beta^\mu=b^\mu+\varpi^{\mu\nu}x_\nu.
\end{equation}
The four-temperature naturally identifies a hydrodynamic frame (defining a four-velocity $u^\mu=T\beta^\mu$, 
being $T=1/\sqrt{\beta^2}$ the proper temperature). The additional Lagrange multiplier $\zeta$ is 
the ratio of the chemical potential and the proper temperature, $\zeta=\mu/T$, and it is constant at global equilibrium.
The tensor $\varpi$ is the thermal vorticity, since from the eq.~\eqref{betakill} one readily
obtains the \eqref{thvort}. The thermal vorticity can be decomposed in two space-like vectors by using 
the four-velocity of the fluid:
\begin{equation*}
\varpi^{\mu\nu}=\epsilon^{\mu\nu\rho\sigma}w_\rho u_\sigma +\alpha^\mu u^\nu - \alpha^\nu u^\mu, 
\end{equation*}
where:
\begin{align}\label{eq: ang vel and acc}
w^\mu=-\frac{1}{2}\epsilon^{\mu\nu\rho\sigma}\varpi_{\nu\rho}u_\sigma, && \alpha^\mu=\varpi^{\mu\nu}u_\nu.
\end{align}
The general expressions of $w^\mu$ and $\alpha^\mu$ can be obtained from the above definition and
the \eqref{thvort}; at global equilibrium they reduce to the ratio of the angular velocity $\omega^\mu$ 
and the four-acceleration $A^\mu$ with the proper temperature $T$. Explicitly, one has:
\be\label{ratiowalpha}
w^\mu=\frac{\omega^\mu}{T}, \qquad \qquad \alpha^\mu=\frac{A^\mu}{T}.
\ee

A method to calculate the expectation values with the density operator \eqref{eq:density op glob eq} 
was proposed in refs.~\cite{Becattini:2020qol,Palermo:2021hlf}. The idea is to analytically 
continue the density operator \eqref{eq:density op glob eq} to imaginary vorticity, 
i.e. setting $\varpi\mapsto -\ii \phi$, and to make an analytic continuation back to real thermal
vorticity \emph{after} the analytic results in $\phi$ are obtained. With this technique,
expectation values are expressed as series of functions and the analytic continuation 
to real thermal vorticity generally requires an intermediate operation
that we dubbed as {\em analytic distillation}. Yet, in the case of the spin density matrix and
the spin polarization vector, we will see that analytic distillation is not necessary and the analytic
continuation can be done straightforwardly.

The expectation value of the quadratic combination of creation and annihilation operators with
the analytically continued density operator reads \cite{Palermo:2021hlf}:
\begin{align}\label{number operator iterative result}
        \langle \wad{p}{s} \wa{p'}{t}\rangle=2\varepsilon' \, \displaystyle \sum_{n=1}^{\infty}(-1)^{2S(n+1)}
    \delta^3(\Lambda^n{\bf{p}}-{\bf{p}'})D(W(\Lambda^n,p))_{ts}
    \e^{-\widetilde{b}\cdot\sum_{k=1}^n\Lambda^kp}\e^{n\zeta}.
\end{align}
In the above series, we have tacitly introduced $\Lambda$ as a notation for the Lorentz 
transformation $\Lambda=\exp\left[-i\phi:J/2\right]$, while $\tilde{b}(\phi)$ is given by:
\begin{equation}\label{eq: def b tilde}
    \tilde{b}^{\mu}(\phi)=\sum_{k=0}^{\infty}\frac{1}{(k+1)!}\phi^{\mu}_{\ \nu_1}
    \phi^{\nu_1}_{\ \nu_2}\dots\phi^{\nu_{k-1}}_{\ \nu_k}b^{\nu_k}.
\end{equation}
The equation \eqref{number operator iterative result}, which applies to both massive and massless particles, 
involves the representation of the transformation $W(\Lambda,p)=[\Lambda p]^{-1}\Lambda [p]$, where $[p]$ 
is the standard Lorentz transformation mapping the conventional basis to particle basis, that is 
$p^\mu=[p]^{\mu}_{\ \nu}\mathfrak{p}^\nu$ as defined in sec. \ref{sec: Polarization QFT}. The transformation
$W(\Lambda,p)$ belongs to the little group of the standard vector $\mathfrak{p}$, that is, it leaves
$\mathfrak{p}$ invariant. For massive spin-S states, $D(W(\Lambda,p))$ is an element of the $S$-irreducible 
representation of the rotation group SO(3), $D(W(\Lambda,p))=D^S(W(\Lambda,p))$, and $W(\Lambda,p)$ is thus 
commonly known as {\em Wigner rotation}. In fact, as a consequence of \eqref{eq s1|p>=0}, in the massless 
case the transformation $W(\Lambda,p)$ is a composition of a rotation with a Lorentz transformation, but
its representation reduces to a phase factor \cite{Weinberg1995qft1}:
\be\label{wigrotmassless}
D(W(\Lambda,p))_{rs}=\exp[\ii \,s \vartheta(\Lambda,p)]\delta_{rs},
\ee
 where $r$ and $s$ can only be $\pm S$, being $S$ the helicity of the particle. The sum over $n$ in eq.
\eqref{number operator iterative result} can be interpreted as the quantum statistics expansion, 
whose first term $n=1$ is the Boltzmann statistics contribution \cite{Becattini:2020qol}.

Using equation \eqref{number operator iterative result}, the particle component of the analytically
continued Wigner function for Dirac fermions appearing in equations \eqref{eq: S in terms of W massive} and 
\eqref{S as function of W} can be obtained \cite{Palermo:2021hlf}:
\begin{equation}\label{eq: exact wigner function}
    W_+(x,k)=\frac{1}{(2\pi)^3}\int \frac{\di^3p}{2\varepsilon} \sum_{n=1}^{\infty}(-1)^{n+1}
\e^{-\tilde{b}(\phi) \cdot \sum_{k=1}^n \Lambda^k p}\e^{n\zeta}\e^{-ix\cdot(\Lambda^np -p)}\e^{n\zeta}
\exp\left(-in\frac{\phi:\Sigma}{2}\right) (m+\slashed{p})\delta^4\left( k-\frac{\Lambda^n p+p}{2} \right),
\end{equation}
where $\slashed{p}=p_\mu\gamma^\mu$, $\gamma^\mu$ being the gamma matrices and $\Sigma^{\mu\nu}=
(i/4)[\gamma^\mu, \gamma^\nu]$ is the generator of Lorentz transformations in the Dirac representation. 
The Wigner function for massless fermions is simply obtained by setting $m=0$ in the eq.~\eqref{eq: exact wigner function}.

Plugging the eq.~\eqref{eq: exact wigner function} in the \eqref{eq: S in terms of W massive}
one can obtain the spin polarization vector of massive Dirac fermions. In view of the relation \eqref{property conservation},
it is possible to compute the integral in eq. \eqref{eq: S in terms of W massive} over a constant-time 
hypersurface $t=t_0$, obtaining:
\begin{align}\label{eq: exact S massive}
S^\mu(p)=&\frac{1}{2m}\frac{\sum_{n=1}^{\infty}(-1)^{n+1} \e^{-\tilde{b}(\phi) \cdot \sum_{k=1}^n 
\Lambda^k p} \e^{n\zeta}
\tr\left(\gamma^\mu\gamma_5\exp [-in\phi:\Sigma/2]\slashed{p}\right)\delta^3(\Lambda^n p -p) }
 { \sum_{n=1}^{\infty}(-1)^{n+1} \e^{-\tilde{b}(\phi) \cdot \sum_{k=1}^n \Lambda^kp}
 \e^{n\zeta}
 \tr\left(\exp [-in\phi:\Sigma/2]\right)\delta^3(\Lambda^n p-p) }.
\end{align}
The above expression is the ratio of two series of $\delta$-functions, which appears daunting. 
This is not surprising, though, as the equation \eqref{eq: S in terms of W massive}, as well as 
\eqref{S as function of W}, was originally derived from the spin density matrix 
\eqref{definition theta field theory}, whose definition is based on $\langle \wad{p}{s} \wa{p}{t}\rangle$.
These expectation values read, according to the equation \eqref{number operator iterative result}:
\begin{align*}
\langle \wad{p}{s} \wa{p}{t}\rangle=2\varepsilon \, \displaystyle 
\sum_{n=1}^{\infty}(-1)^{2S(n+1)} \delta^3(\Lambda^n{\bf p}-{\bf p}) D^S(W(\Lambda^n,p))_{rs}
    \e^{-\widetilde{b}\cdot\sum_{k=1}^n\Lambda^kp} \e^{n\zeta},
\end{align*}
and they vanish unless $\Lambda^np=p$. This relation is fulfilled for specific $n\neq 1$ or if 
$\Lambda p=p$. In the latter case, the equation holds $\forall n$, whereas if the equation is solved 
for $n\neq 1$ and given $p$, $\Lambda$ would be a discrete transformation, which would make the analytic 
continuation impossible; therefore, we will focus on the case $\Lambda p = p$.
This constraint requires $\Lambda$ to belong to the little group of $p$, i.e. the group of 
transformations leaving $p$ invariant. By using the exponential parametrization of the Lorentz group, 
$\Lambda=\exp[-i\phi:J/2]$, and expanding for infinitesimal $\phi$: 
\begin{equation}\label{eq: phi p = 0}
  \exp[-i\phi:J/2] p = p \implies \phi^{\mu\nu} p_\nu=0,
\end{equation} 
being $(J_{\mu\nu})^\rho_\sigma = \ii (\delta^\rho_\mu g_{\nu\sigma} - \delta^\rho_\nu g_{\mu\sigma})$.
The general solution of eq. \eqref{eq: phi p = 0} can be expressed in terms of an auxiliary vector 
$\xi^\mu$:
\begin{equation}\label{eq:constraint global pol}
 \xi^\rho=-\frac{1}{2m}\epsilon^{\rho\mu\nu\sigma}\phi_{\mu\nu}p_\sigma.
\end{equation}
In this case, it can be shown that:
\begin{align}\label{eq: phi:phi}
   \phi^{\mu\nu} = \epsilon^{\mu\nu\rho\sigma}\xi_\rho \frac{p_\sigma}{m}, && 
   \phi:\phi = -2\xi^2, && \phi:\widetilde{\phi}=0,
\end{align}
where the tensor $\tilde{\phi}$ is the dual of $\phi$: 
$$
\tilde{\phi}^{\mu\nu}=\frac{1}{2}\epsilon^{\mu\nu\rho\sigma}\phi_{\rho\sigma}.
$$
Notice that, since $\xi\cdot p =0$ and $p$ is time-like, $\xi$ is a space-like vector and $-\xi^2>0$. 
If $\Lambda p = p$, it turns out that all the $\delta$ functions in the equation \eqref{eq: exact S massive}
reduce to $\delta^3(0)$ that simplifies in the ratio, leaving:
\begin{align}\label{exactSmassive2}
S^\mu(p)=&\frac{1}{2m}\frac{\sum_{n=1}^{\infty}(-1)^{n+1} \e^{-\tilde{b}(\phi) \cdot \sum_{k=1}^n \Lambda^k p} \e^{n\zeta}
     \tr\left(\gamma^\mu\gamma_5\exp [-in\phi:\Sigma/2]\slashed{p}\right)\delta^3(0) }
    { \sum_{n=1}^{\infty}(-1)^{n+1} \e^{-\tilde{b}(\phi) \cdot \sum_{k=1}^n \Lambda^k p}\e^{n\zeta} 
    \tr\left(\exp [-in\phi:\Sigma/2]\right)\delta^3(0) } \nonumber \\
    =&\frac{1}{2m}\frac{\sum_{n=1}^{\infty}(-1)^{n+1} \e^{-\tilde{b}(\phi) \cdot \sum_{k=1}^n \Lambda^k p}\e^{n\zeta} 
    p_\nu \tr\left(\gamma^\mu\gamma_5\exp [-in\phi:\Sigma/2]\gamma^\nu\right) }
    { \sum_{n=1}^{\infty}(-1)^{n+1} \e^{-\tilde{b}(\phi) \cdot \sum_{k=1}^n \Lambda^k p} \e^{n\zeta}
    \tr\left(\exp [-in\phi:\Sigma/2]\right) }.
\end{align}
This formula coincides with eq. (9.6) reported in \cite{Palermo:2021hlf}, although its derivation 
was not carried out in detail. The constraint $\Lambda p = p$ implies more simplifications, particularly:
\begin{equation}\label{eq:simplification tbp}
\tilde{b}(\phi) \cdot \sum_{k=1}^n \Lambda^k p = n\tilde{b}(\phi)\cdot p=n\sum_{k=0}^{\infty}
\frac{1}{(k+1)!}p_\mu (\underbrace{\phi^\mu_{\ \alpha_1}\phi^{\alpha_1}_{\ \alpha_2}\dots
\phi^{\alpha_{k-1}}_{\ \ \ \alpha_k}}_{k \text{ times}})b^{\alpha_k}=nb\cdot p,
\end{equation}
where the definition \eqref{eq: def b tilde} and eq. \eqref{eq: phi p = 0} have been used. Thanks to 
equations \eqref{eq: phi:phi}, especially $\phi:\widetilde{\phi}=0$, also the traces in the equation
\eqref{exactSmassive2} take a simpler form (calculations are reported in appendix \ref{app: traces}):
\begin{subequations}\label{eq: identities traces}
    \begin{align}
        \tr\left(\exp [-in\phi:\Sigma/2]\right)=&4\cos\left(\frac{n}{2}\sqrt{\frac{\phi:\phi}{2}}\right),\\
        \tr(\gamma^\nu \gamma^\mu \exp [-in\phi:\Sigma/2])=&4g^{\mu\nu}\cos\left(\frac{n}{2}
        \sqrt{\frac{\phi:\phi}{2}}\right)+4\phi^{\mu\nu}\frac{\sin\left(\frac{n}{2}
        \sqrt{\frac{\phi:\phi}{2}}\right)}{\sqrt{\frac{\phi:\phi}{2}}},\\
        \tr(\gamma^\nu \gamma^\mu \gamma_5 \exp [-in\phi:\Sigma/2])=&4i\tilde{\phi}^{\mu\nu}
        \frac{\sin\left(\frac{n}{2}\sqrt{\frac{\phi:\phi}{2}}\right)}{\sqrt{\frac{\phi:\phi}{2}}}.
    \end{align}
\end{subequations}
By plugging the above traces and the eq.~\eqref{eq:simplification tbp} into the \eqref{exactSmassive2}
we obtain:
$$
S^\mu(p)=
\frac{i\tilde{\phi}^{\mu\nu}p_\nu}{2m\sqrt{-\xi^2}}\frac{\sum_{n=1}^{\infty}(-1)^{n+1} \e^{-nb \cdot p+n\zeta}
\sin\left(n\sqrt{-\xi^2}/2\right)}{\sum_{n=1}^{\infty}(-1)^{n+1} \e^{-nb \cdot p+n\zeta}
\cos\left(n\sqrt{-\xi^2}/2\right)}.
$$
Both series in the above equation are convergent $\forall \phi\in \mathbb{R}$ and $b\cdot p >\zeta$. 
Expanding $\tilde{\phi}$, the summation yields:
\begin{equation}\label{eq: intermediate S with theta}
S^\mu(p)=\frac{i\epsilon^{\mu\nu\rho\sigma}\phi_{\rho\sigma}p_\nu}
{4m\sqrt{-\xi^2}}\frac{\sin\left(\sqrt{-\xi^2}/2\right)}
{\cos\left(\sqrt{-\xi^2}/2\right)+\e^{-b\cdot p +\zeta}},
\end{equation}
which is an analytic function in $\phi$.

The equation \eqref{eq: intermediate S with theta} can be analytically continued to real thermal vorticity. 
Introducing the vector:
\begin{equation}\label{eq: definition theta vector}
    \theta^\mu =-\frac{1}{2m}\epsilon^{\mu\nu\rho\sigma}\varpi_{\nu\rho}p_\sigma,
\end{equation} 
the mapping $\phi\mapsto i \varpi$ implies $\xi\mapsto i\theta$. Notice that, the vector $\theta^\mu$ is just
the ratio between the local angular velocity seen by the particle ($p/m$ replacing $u$ in eq. 
\eqref{eq: ang vel and acc}) and the temperature. Therefore, the continuation of equation 
\eqref{eq: intermediate S with theta} reads:
\begin{equation}\label{spinvectancon}
S^\mu(p)=-\frac{1}{4m}\epsilon^{\mu\nu\rho\sigma}\varpi_{\nu\rho}p_\sigma
\;\frac{1}{\sqrt{-\theta^2}}\frac{\sinh\left(\sqrt{-\theta^2}/2\right)}
{\cosh\left(\sqrt{-\theta^2}/2\right)+\e^{-b\cdot p +\zeta}}.
\end{equation}
The polarization vector can also be expressed solely in terms of $\theta^\mu$, yielding a 
more suggestive expression:
\begin{equation}\label{eq:spin exact with s}
S^\mu(p)=\frac{1}{2}
\;\frac{\theta^\mu}{\sqrt{-\theta^2}}\frac{\sinh\left(\sqrt{-\theta^2}/2\right)}
{\cosh\left(\sqrt{-\theta^2}/2\right)+\e^{-b\cdot p +\zeta}}=\frac{1}{2}
\;\hat\theta^\mu P_{1/2}\left(\sqrt{-\theta^2},b\cdot p-\zeta\right).
\end{equation}
The above equation has been written such that the factor $1/2$ is the spin of the particle, 
the vector $\hat{\theta}^\mu=\theta^\mu/\sqrt{-\theta^2}$ provides the direction of $S^\mu(p)$, 
while its magnitude is determined by the weight function: 
\begin{equation}\label{eq: S1/2 distribution}
P_{1/2}(x,y)=\frac{\sinh\left(x/2\right)}{\cosh\left(x/2\right) +\e^{-y}}.
\end{equation}
where $x= \sqrt{-\theta^2}$ and $y= b\cdot p - \zeta$. Note that, since the polarization 
vector of a spin-S particle is defined as:
$$
P^\mu=\frac{S^\mu}{S},
$$
the function $P_{1/2}$ is in fact the polarization itself. The function $P_{1/2}$ monotonically 
increases in both arguments for $x\geq 0$, and it is bounded in the interval $0\leq P_{1/2}\leq 1$ 
at it should. Its limiting values are:
\begin{equation}\label{eq:asymptotics S}
 \lim_{x\rightarrow\infty}P_{1/2}(x,y)=1, \qquad \qquad 
 P_{1/2}(x,y) \simeq \frac{x}{2}(1-n_F(y)) \qquad {\rm for} \;\; x \ll 1
\end{equation}
with $n_F(y)=[\exp(y)+1]^{-1}$ as in \eqref{eq: fermi-dirac}. Therefore:
\begin{subequations}
\begin{align}
\lim_{\sqrt{-\theta^2}\rightarrow \infty}S^\mu(p)&=\frac{1}{2}\, \frac{\theta^\mu}{\sqrt{-\theta^2}},\label{eq: pol bound}\\
S^\mu(p)&\simeq\frac{\theta^\mu}{4}(1-n_F(b\cdot p-\zeta)),
\qquad \qquad {\rm if} \;\;
\sqrt{-\theta^2}\ll 1.\label{eq: small vorticity massive}
\end{align}
\end{subequations}
Recalling the definition of $\theta^\mu$, eq. \eqref{eq: definition theta vector}, it can be 
seen that the second limit agrees with the equation
\eqref{eq:linear spin vector vort} in the case of global equilibrium. 
Besides, equation \eqref{eq: pol bound} shows that, for an infinitely large vorticity, particles
become fully polarized in the direction of $\theta$.

It is worth discussing in more detail the obtained results. A crucial role in the determination 
of the exact formula of the spin polarization vector has been played by the constraint 
$\Lambda p = p$, equivalent to $\phi_{\mu\nu} p^\nu = 0$, which has been used to determine the 
form of the imaginary vorticity $\phi_{\mu\nu}$ in the eq.~\eqref{eq: phi:phi}. 
Nevertheless, it should be emphasized that, after the analytic continuation $\phi = \ii \varpi$ 
of the equation \eqref{eq: intermediate S with theta} to the final \eqref{spinvectancon}, such constraint 
does not extend to $\varpi$. Otherwise stated, the continuation of $\phi$ to real thermal vorticity
does not bring the constraint along.

It should also be emphasized that our obtained expression of the spin polarization vector has the
correct bound of polarization, given by the equation \eqref{eq: pol bound}. This is in 
important point, as a violation of the unitarity bound of polarization was noted in ref.~\cite{Florkowski:2018fap} 
where the spin polarization vector was calculated at all orders in thermal vorticity in the 
Boltzmann limit. The violation was attributed to a problem in the {\em ansatz} of the Wigner 
function used in the derivation. Indeed, it can be shown that the violation also appears 
in the single quantum relativistic particle framework \cite{becattini2020polarization} if the constraint 
$\phi_{\mu\nu} p^\nu = 0$ is neglected before the analytic continuation to real thermal vorticity.

\subsection{Massless Dirac fermions}

We can now move to the case of the massless Dirac field. Using the eq. \eqref{S as function of W} and 
the exact Wigner function of Dirac fermions \eqref{eq: exact wigner function} with $m=0$ and $\zeta=0$, and calculating
the integrals in eq.~\eqref{S as function of W} over a constant-time hypersurface, one finds:
\begin{equation}\label{eq:exac PL massless} 
\Pi^\mu = \frac{p^\mu}{2} \frac{\sum_{n=1}^{\infty}(-1)^{n+1} \e^{-\tilde{b}(\phi) \cdot \sum_{k=1}^n \Lambda^k p} 
     \tr\left(\slashed{q}\gamma_5\exp [-in\phi:\Sigma/2]\slashed{p}\right)\delta^3(\Lambda^n p -p) }
    { \sum_{n=1}^{\infty}(-1)^{n+1} \e^{-\tilde{b}(\phi) \cdot \sum_{k=1}^n \Lambda^k p} 
    \tr\left(\slashed{q}\exp [-in\phi:\Sigma/2]\slashed{p}\right)\delta^3(\Lambda^n p-p) }.     
\end{equation}
The constraint $\Lambda p=p$ implies $\phi^{\mu\nu} p_\nu=0$, just like in the massive case.  
Indeed, as it is shown in appendix \ref{app: constraint}, the most general decomposition of an anti-symmetric 
tensor solving eq. \eqref{eq: phi p = 0} for a light-like $p$ is:
\begin{equation}\label{eq:constrained vort massless}
 \phi^{\mu\nu}=\epsilon^{\mu\nu\rho\sigma}\frac{h_\rho p_\sigma}{p\cdot q},\qquad \qquad h^\mu =
 -\frac{1}{2}\epsilon^{\mu\nu\rho\sigma}\phi_{\nu\rho}q_\sigma.
\end{equation}
with $h \cdot q = 0$. All the traces in the eq.~\eqref{eq:exac PL massless} can be simplified much like 
in the massive case. Indeed, it can be shown that:
\begin{align}\label{eq: phi:phi massless}
    \phi:\phi=2\eta^2, && \phi:\tilde{\phi}=0,
\end{align}
where we have defined:
\begin{equation}\label{def: eta massless}
\eta=\frac{h\cdot p}{q\cdot p}=\frac{\widetilde{\phi}^{\mu\nu} q_\mu p_\nu}{p\cdot q}=\frac{1}{2(p\cdot q)}
\epsilon^{\mu\nu\alpha\beta}\phi_{\alpha\beta}p_\nu q_\mu.
\end{equation}
Since $\phi:\tilde{\phi}=0$, the identities \eqref{eq: identities traces} hold, and taking into account 
the eq.~\eqref{eq:simplification tbp} we have:
\begin{equation*}
\Pi^\mu(p)=
\frac{p^\mu}{2\abs{\eta}}\frac{i\tilde{\phi}^{\alpha\beta}q_\alpha p_\beta}{p\cdot q}\frac{\sum_{n=1}^{\infty}(-1)^{n+1} 
\e^{-nb \cdot p} \sin\left(n\abs{\eta}/2\right)}{\sum_{n=1}^{\infty}(-1)^{n+1} \e^{-nb \cdot p}
\cos\left(n\abs{\eta}/2\right)}.
\end{equation*}
The above series converges for $b\cdot p>0$, and the summation yields:
\begin{equation*}
\Pi^\mu(p)=\frac{p^\mu}{2} \frac{i\eta}{\abs{\eta}}\frac{\sin\left(\abs{\eta}/2\right)}
{\cos\left(\abs{\eta}/2\right)+\e^{-b\cdot p}}=\frac{p^\mu}{2}\frac{i\sin\left(\eta/2\right)}
{\cos\left(\eta/2\right)+\e^{-b\cdot p}},
\end{equation*}
where in the last step we have used the parity of the trigonometric functions.

The latter result can be readily continued to real thermal vorticity. Introducing:
\begin{equation}
    H=\frac{\tilde{\varpi}^{\mu\nu}q_\mu p_\nu}{p\cdot q}=\frac{1}{2(p\cdot q)}
    \epsilon^{\mu\nu\alpha\beta}\varpi_{\alpha\beta}p_\nu q_\mu,
\end{equation}
and realizing that $\phi\mapsto i\varpi$ implies $\eta\mapsto iH$, one finds:
\begin{equation}\label{eq:exact pol massless 1/2}
\Pi^\mu(p)=-\frac{p^\mu}{2}\frac{\sinh\left(H/2\right)}{\cosh\left(H/2\right)+\e^{-b\cdot p}}.
\end{equation}
The Lorentz scalar $H$ can be written in a way which is independent of $q$ by breaking
manifest covariance. Since $\mathfrak{q}=(\kappa,0,0,-\kappa)$ is the parity conjugate of $\mathfrak{p}$ 
and $[p]^\mu_{\ \nu}\mathfrak{p}^\nu=p^\mu=(\varepsilon,{\bf p})$, where $\varepsilon=\|{\bf p}\|$ is 
the energy, it can be readily shown that:
$$
q^\mu = [p]^{\mu}_{\ \nu}\mathfrak{q}^\nu=\kappa^2 (1/\varepsilon,-{\bf p}/\varepsilon^2).
$$
Therefore:
\begin{equation*}
     H = \frac{1}{2(p\cdot q)}\epsilon^{\mu\nu\alpha\beta}\varpi_{\alpha\beta}p_\nu q_\mu
     =  \frac{1}{4\kappa^2}\epsilon^{0 n \alpha\beta} \varpi_{\alpha\beta} p_n q_0
     + \frac{1}{4\kappa^2}\epsilon^{m 0 \alpha\beta} \varpi_{\alpha\beta} p_0 q_m
     =  \frac{1}{2 \varepsilon}\epsilon^{0 n \alpha\beta} \varpi_{\alpha\beta} p_n,
\end{equation*}
for the combinations where $p$ and $q$ both have a space index vanish due to 
$\epsilon^{m n\alpha \beta}p_n q_m=-\epsilon^{m n\alpha \beta}p_n p_m=0$. Defining:
\begin{align*}
    \varphi^n=-\frac{1}{2}\epsilon^{n\alpha\beta0}\varpi_{\alpha\beta},
     && \hat{\bf{p}}=\frac{\bf{p}}{\varepsilon}.
\end{align*}
we obtain:
\begin{align}\label{eq: H as phi*p}
    H &=-\frac{1}{2 \varepsilon}\epsilon^{0 n \alpha\beta} \varpi_{\alpha\beta} p_n=
     \frac{1}{\varepsilon} \varphi^n p_n = -\bm{\varphi}\cdot \hat{\bf{p}} .
\end{align}

We can finally rewrite the PL vector for massless particles using $\bf{\varphi}$ as: 
\begin{equation*}
\Pi^\mu(p)=\frac{p^\mu}{2}P_{1/2}\left(\bm{\varphi}\cdot {\bf \hat{p}},b\cdot p\right),
\end{equation*}
where we made use of the function $P_{1/2}$ defined in eq. \eqref{eq: S1/2 distribution}.
The limits of very large and very small thermal vorticity, are easily obtained from 
\eqref{eq:asymptotics S}:
\begin{align*}
\lim_{\abs{\bm{\varphi}\cdot{\bf\hat{p}}}\rightarrow \infty}\Pi^\mu(p)\simeq&\frac{p^\mu}{2}{\rm sgn}
(\bm{\varphi}\cdot \bf{\hat{p}}), \\
\Pi^\mu(p) \simeq &\frac{p^\mu}{4} \bm{\varphi} \cdot {\bf\hat{p}} (1-n_F(b\cdot p)), \qquad \qquad
{\rm if} \;\; \abs{\bm{\varphi}\cdot {\bf\hat{p}}} \ll 1.
\end{align*}
%

\section{Particles with any spin}
\label{sec: exact any spin}

The previous results can be extended to particles of any spin $S$. For this purpose, since we are 
working at global equilibrium, the spin density matrix can be used in the first place without 
introducing the covariant Wigner function. Still, the calculation requires the analytic continuation 
of the density operator to imaginary thermal vorticity. Using the definition 
\eqref{definition theta field theory} and the exact expression of $\langle\wad{p}{s}\wa{p'}{r}\rangle$ at 
general global equilibrium, equation \eqref{number operator iterative result}, one finds:
\begin{align*}
\Theta_{rs}(p)=\frac{\langle \wad{p}{s}\wa{p}{r}\rangle}{\sum_t\langle \wad{p}{t}\wa{p}{t}\rangle}
=\frac{\sum_{n=1}^\infty (-1)^{2S(n+1)} 
\e^{-\widetilde{b}\cdot\sum_{k=1}^n\Lambda^kp}\e^{n\zeta}D_{rs}(W(\Lambda^n,p))\delta^3(\Lambda^np-p)}
{\sum_{n=1}^\infty (-1)^{2S(n+1)} \e^{-\widetilde{b}\cdot
\sum_{k=1}^n\Lambda^kp}\e^{n\zeta}\tr[D(W(\Lambda^n,p))]\delta^3(\Lambda^np-p)}.
\end{align*}
According to the discussion in section \ref{sec: exact pol fermions}, the Dirac $\delta$-function 
constrains the tensor $\phi$ such that the equation $\Lambda p = p$ is fulfilled, hence $\phi_{\mu\nu}p^\nu=0$. 
Therefore all the Dirac $\delta$-functions boil down to a common divergent factor $\delta^3(0)$ in 
the numerator and the denominator, similarly to eq. \eqref{exactSmassive2}, and so:
 \begin{align}\label{eq:exact spin density any spin}
\Theta_{rs}(p)=\frac{\sum_{n=1}^\infty (-1)^{2S(n+1)} 
\e^{-nb\cdot p}\e^{n\zeta}D_{rs}(W(\Lambda^n,p))}
{\sum_{n=1}^\infty (-1)^{2S(n+1)} \e^{-nb\cdot
 p}\e^{n\zeta}\tr[D(W(\Lambda^n,p))]},
\end{align}
where we have set $-\widetilde{b}\cdot\sum_{k=1}^n\Lambda^kp=nb\cdot p$ according to the 
eq.~ \eqref{eq:simplification tbp}.
Using the above equation, one can determine the exact form at global equilibrium of spin related
observables, such as the polarization vector or the spin alignment parameter, for both massive 
and massless particles of any spin. In what follows, we will confine ourselves with the spin vector,
tackling the massive and massless case separately.

\subsection{Massive particles}

The first step to derive the spin density matrix and the spin polarization vector, both for massive 
and massless particles, is to determine the Wigner rotation for Lorentz transformations $\Lambda$ in 
the little group of $p$; this calculation is carried out in the appendix \ref{app:wigner phase}. 

For massive particles, this Wigner rotation turns out to be:
\begin{equation}  \label{eq: wigner rot massiva simplified}
    D^{S}(W(\Lambda,p))=\exp\left[-\ii\bm{\xi}_0\cdot D^S(\textbf{J})\right] \; ,
\end{equation}
where $\xi_0^\mu=[p]^{-1\mu}_{\; \nu}\xi^\nu = (0,{\bm \xi}_0)$ and $\textbf{J}$ is the three-vector 
of the generators of SO(3). The time-component of $\xi_0$ vanishes because $\xi_0 \cdot \mathfrak{p} = 
\xi \cdot p = 0$. The equation \eqref{eq: wigner rot massiva simplified} shows that the Wigner rotation 
corresponds to a rotation of an angle $\sqrt{\bm{\xi}_0\cdot\bm{\xi}_0} =\sqrt{-\xi \cdot \xi}$ around 
the axis $\bm{\hat{\xi}_0}=\bm{\xi}_0/\sqrt{\bm{\xi}_0\cdot\bm{\xi}_0}$. Using this result, the 
eq. \eqref{eq:exact spin density any spin} reduces to:
\begin{align*}
\Theta(p)=\frac{\sum_{n=1}^\infty (-1)^{2S(n+1)} \e^{-nb \cdot p}\e^{n\zeta}\e^{-\ii n\bm{\xi}_0\cdot D^S(\textbf{J})}}
{\sum_{n=1}^\infty (-1)^{2S(n+1)} \e^{-nb\cdot p}\e^{n\zeta}\tr\left(\e^{-\ii n\bm{\xi}_0\cdot D^S(\textbf{J})}\right)}.
\end{align*}
The spin polarization vector can be readily found from the above equation using the eq.~\eqref{eq:spin vect tr J}:
\begin{align}\label{genspinpol}
S^\mu(p)=\sum_{i=1}^3 [p]^\mu_{\ i}\tr\left(\Theta(p)D^S(\mathrm{J}^i)\right)=\sum_{i=1}^3 [p]^\mu_{\ i}
\frac{\sum_{n=1}^\infty (-1)^{2S(n+1)} \e^{-nb \cdot p}\e^{n\zeta}\tr\left(\e^{-\ii n\bm{\xi}_0\cdot D^S(\textbf{J})}
D^S({\rm J}^i)\right)} {\sum_{n=1}^\infty (-1)^{2S(n+1)} 
\e^{-nb\cdot p}\e^{n\zeta}\tr\left(\e^{-\ii n\bm{\xi}_0\cdot D^S(\textbf{J})} \right)}.
\end{align}
Developing the expression \eqref{genspinpol} requires some intermediate steps. We can calculate the trace 
in the denominator by choosing the $z$ axis along $\bm{\xi}_0$ so that $D^S(\rm{J}^3)$ is diagonal:
\begin{equation}\label{eq: character}
 \tr\left(\e^{-\ii n \bm{\xi_0}\cdot D^S(\bf{J})}\right)=
 \sum_{k=-S}^{S} \e^{-\ii k\,n\sqrt{-\xi^2}} \equiv \chi_n^S\left(\sqrt{-\xi^2}\right).
\end{equation}
Besides:
\begin{align*}
\tr\left(\e^{-\ii n \bm{\xi}_0\cdot D^S(\textbf{J}) } D^S(\mathrm{J}^i)\right)
=\frac{\ii} {n}\frac{\partial}{\partial \xi_0^i} \tr\left(\e^{-\ii n \bm{\xi}_0\cdot 
D^S(\textbf{J}) }\right)=\frac{\ii}{n}\frac{{{\xi_0}^i}}{\sqrt{-\xi^2}}\frac{\partial\chi_n^S}
{\partial \sqrt{-\xi^2}},
\end{align*}
where we used $\bm{\xi}_0^2=-\xi^2$, that follows from Lorentz invariance.
The standard boost in eq. \eqref{eq:spin vect tr J} is such that $\sum_{i=1}^3[p]^\mu_{\ i}
{\xi_0}^i=\xi^\mu$, hence the polarization vector \eqref{genspinpol} becomes:
\begin{align}\label{eq: exact spin character}
    S^\mu(p)=\frac{i\xi^\mu}{\sqrt{-\xi^2}}\frac{\sum_{n=1}^{\infty}(-1)^{2S(n+1)}(1/n) \e^{-nb\cdot p}\e^{n\zeta}
    \frac{\partial\chi^S_n}{\partial \sqrt{-\xi^2}}}{\sum_{n=1}^{\infty}(-1)^{2S(n+1)}\e^{-nb\cdot p}\e^{n\zeta}
    \chi_n^S\left(\sqrt{-\xi^2}\right)}.
\end{align}
The series can be rewritten using the eq. \eqref{eq: character}: 
\begin{align*}
    &\sum_{n=1}^{\infty}(-1)^{2S(n+1)}\e^{-nb\cdot p}\e^{n\zeta}\chi_n^S\left(\sqrt{-\xi^2}\right)=
    \sum_{n=1}^{\infty}\sum_{k=-S}^{S}(-1)^{2S(n+1)}\e^{-nb\cdot p+n\zeta-ik\,n\sqrt{-\xi^2}}=
    \sum_{k=-S}^{S}\frac{1}{\e^{b\cdot p-\zeta + ik\sqrt{-\xi^2}}-(-1)^{2S}},\\
    &\sum_{n=1}^{\infty}(-1)^{2S(n+1)}\e^{-nb\cdot p}\e^{n\zeta}\frac{1}{n}\frac{\partial\chi^S_n}{\partial \sqrt{-\xi^2}}=
    \sum_{n=1}^{\infty}(-1)^{2S(n+1)}\e^{-nb\cdot p + n\zeta}
    \sum_{k=-S}^{S}-ik\e^{-ik\,n\sqrt{-\xi^2}}=\sum_{k=-S}^{S}\frac{-ik}{\e^{b\cdot p-\zeta + ik\sqrt{-\xi^2}}-(-1)^{2S}},
\end{align*}
where, in both cases, the finite sum over $k$ was exchanged with the series in $n$.
 
After having resummed the series in $n$, the analytic continuation is readily done by just mapping 
$\phi\mapsto i\varpi$ and $\xi \mapsto i\theta$. Therefore, the spin polarization vector for a massive 
spin-S particle and for real thermal vorticity reads:
\begin{equation}\label{eq: spin sum of statistics}
    S^\mu(p)=\frac{\theta^\mu}{\sqrt{-\theta^2}}\frac{\sum_{k=-S}^{S} k  
    \left[ \e^{b\cdot p-\zeta - k\sqrt{-\theta^2}}-(-1)^{2S} \right]^{-1}}
    {\sum_{k=-S}^{S} \left[ \e^{b\cdot p-\zeta - k\sqrt{-\theta^2}}-(-1)^{2S} \right]^{-1}}.
\end{equation}
The above formula shows that the spin polarization vector at global equilibrium can be expressed 
as a finite sum of Fermi-Dirac or Bose-Einstein distributions functions, depending on the spin, 
where $\sqrt{-\theta^2}$ acts as a sort of chemical potential. It can be checked that this expression 
reduces to \eqref{eq: exact S massive} for $S=1/2$. 
We can also write the spin polarization vector extending eq. \eqref{eq:spin exact with s} and introducing the polarization function $P_{S}$ for spin-S fields:
\begin{align}\label{eq:spin distrib any spin}
    S^\mu(p)=S\frac{\theta^\mu}{\sqrt{-\theta^2}}P_{S}\left(\sqrt{-\theta^2},b\cdot p-\zeta\right), 
    &&P_{S}\left(\sqrt{-\theta^2},b\cdot p-\zeta\right)=\frac{1}{S}\frac{\sum_{k=-S}^{S}k 
    \left[\e^{b\cdot p-\zeta -k\sqrt{-\theta^2}}-(-1)^{2S}\right]^{-1}}{\sum_{k=-S}^{S} 
    \left[ \e^{b\cdot p -\zeta- k\sqrt{-\theta^2}}-(-1)^{2S} \right]^{-1}}.
\end{align}

In the limit of small $\sqrt{-\theta^2}$, the eq. \eqref{eq: spin sum of statistics} can
be approximated by:
\begin{equation*}
S^\mu(p)=\theta^\mu(1+(-1)^{2S}n_{F/B}(b\cdot p-\zeta))\frac{\sum_{k=-S}^{S}k^2}
{\sum_{k=-S}^{S}1}=\theta^\mu\frac{S(S+1)}{3}(1+(-1)^{2S}n_{F/B}(b\cdot p-\zeta)),
\end{equation*}
with $n_F$ being the Fermi-Dirac distribution, eq. \eqref{eq: fermi-dirac}, and 
$n_B(b\cdot p-\zeta)=[\exp(b\cdot p-\zeta)-1]^{-1}$ is the Bose-Einstein distribution. 
The above result is in agreement with the known linear approximations in thermal vorticity 
\cite{Becattini:2016gvu}.

We can also check the limit of Boltzmann statistics. Since the sum over $n$ corresponds to 
the quantum statistics expansion, the Boltzmann case is obtained retaining the $n=1$ term
in the series. It is straightforward to see that, after mapping $\xi\mapsto i\theta$, 
eq. \eqref{eq: exact spin character} yields:
\begin{equation*}
 S^\mu(p)=\frac{\theta^\mu}{\sqrt{-\theta^2}}\frac{\chi'_1\left(\sqrt{-\theta^2}\right)}
 {\chi_1\left(\sqrt{-\theta^2}\right)},
\end{equation*}
where with $\chi'$ we denote the derivative of $\chi$ with respect to $\sqrt{-\theta^2}$.
This expression is formally the same as that in refs. \cite{Becattini:2007nd,Becattini:2009wh}
obtained for a rotating fluid.

\subsection{Massless particles}

In the massless case, due to eq. \eqref{eq s1|p>=0}, the ``Wigner rotation" 
appearing in \eqref{eq:exact spin density any spin} is just the eq.~\eqref{wigrotmassless}:
$$
    D(W(\Lambda,p))_{hk}=\e^{i\vartheta(\Lambda,p) h}\delta_{hk},
$$
where $h,k$ can only be $+S$ or $-S$. The derivation of the angle $\vartheta$ associated to a 
transformation $\Lambda$ such that $\Lambda p=p$ is reported in appendix \ref{app:wigner phase};
the result is:
$$
\vartheta(\Lambda,p)=\frac{h\cdot p}{q\cdot p} = \eta
$$
where $\eta$ was defined in eq. \eqref{def: eta massless}.

Now we can obtain the spin density matrix and the spin polarization vector. From eq. 
\eqref{eq:exact spin density any spin} one has:
\begin{equation}
    \Theta_{hk}(p)=\frac{\sum_{n=1}^\infty (-1)^{2S(n+1)} 
    \e^{-nb\cdot p}\e^{\ii n \eta h}\delta_{hk}}{\sum_{n=1}^\infty (-1)^{2S(n+1)} \e^{-n b\cdot p}2\cos n\eta},
\end{equation}
whence, by using the equation \eqref{polvectmass0}, the spin polarization vector
is obtained:
\begin{align*}
\Pi^\mu(p)=p^\mu \sum_{h=\pm S}h \Theta_{hh}=ip^\mu S\frac{\sum_{n=1}^\infty (-1)^{2S(n+1)}
 \e^{-n b\cdot p}\sin\left(n\eta\,S\right)}{\sum_{n=1}^\infty (-1)^{2S(n+1)}
 \e^{-n b\cdot p}\cos\left(n\eta\,S\right)},
\end{align*}
where it has been taken into account that $h,k$ can only take on values $\pm S$, with $S=1/2,1,3/2\dots$ 
denoting the magnitude of the helicity of the particle. The above series can be straightforwardly resummed, 
yielding:
\begin{equation*}
\Pi^\mu(p)=ip^\mu S\frac{\sin\left(\eta\,S\right)}{\cos\left(\eta\,S\right)-(-1)^{2S}\e^{-b\cdot p}}.
\end{equation*}
This result can be continued to real thermal vorticity. Mapping $\phi\mapsto i\varpi$ and $\eta\mapsto i H$ 
we have:
\begin{equation}\label{finalspinmasszero}
\Pi^\mu(p)=-p^\mu S\frac{\sinh\left(S\,H\right)}{\cosh\left(S\,H\right)-(-1)^{2S}\e^{-b\cdot p}}
=p^\mu S\frac{\sinh\left(S\,\bm{\varphi}\cdot \hat{{\bf p}}\right)}
{\cosh\left(S\,\bm{\varphi}\cdot \hat{\bf p}\right)-(-1)^{2S}\e^{-b\cdot p}},
\end{equation}
where in the last step we have used equation \eqref{eq: H as phi*p}. The equation \eqref{finalspinmasszero}
reproduces equation \eqref{eq:exact pol massless 1/2} for $S=1/2$.

\section{Application to $\Lambda$ polarization in heavy ion collisions}
\label{sec: phenomenology}

In relativistic heavy ion collisions, the theoretical estimates of the spin polarization vector 
of spin 1/2 hyperons are obtained at linear order in thermal vorticity, see eq. \eqref{eq:linear spin vector vort}. 
Even though it is known that thermal vorticity is generally $\ll 1$ throughout the freeze-out hypersurface 
\cite{Becattini:2021iol} at high energy, it would be important to have a quantitative assessment the accuracy 
of this approximation. Indeed, we are in a position to provide such an assessment by comparing the exact 
formula of the spin polarization vector at global equilibrium - that is with constant thermal vorticity -
with its linear approximation.

\begin{figure}
    \centering
    \includegraphics[width=0.49\textwidth]{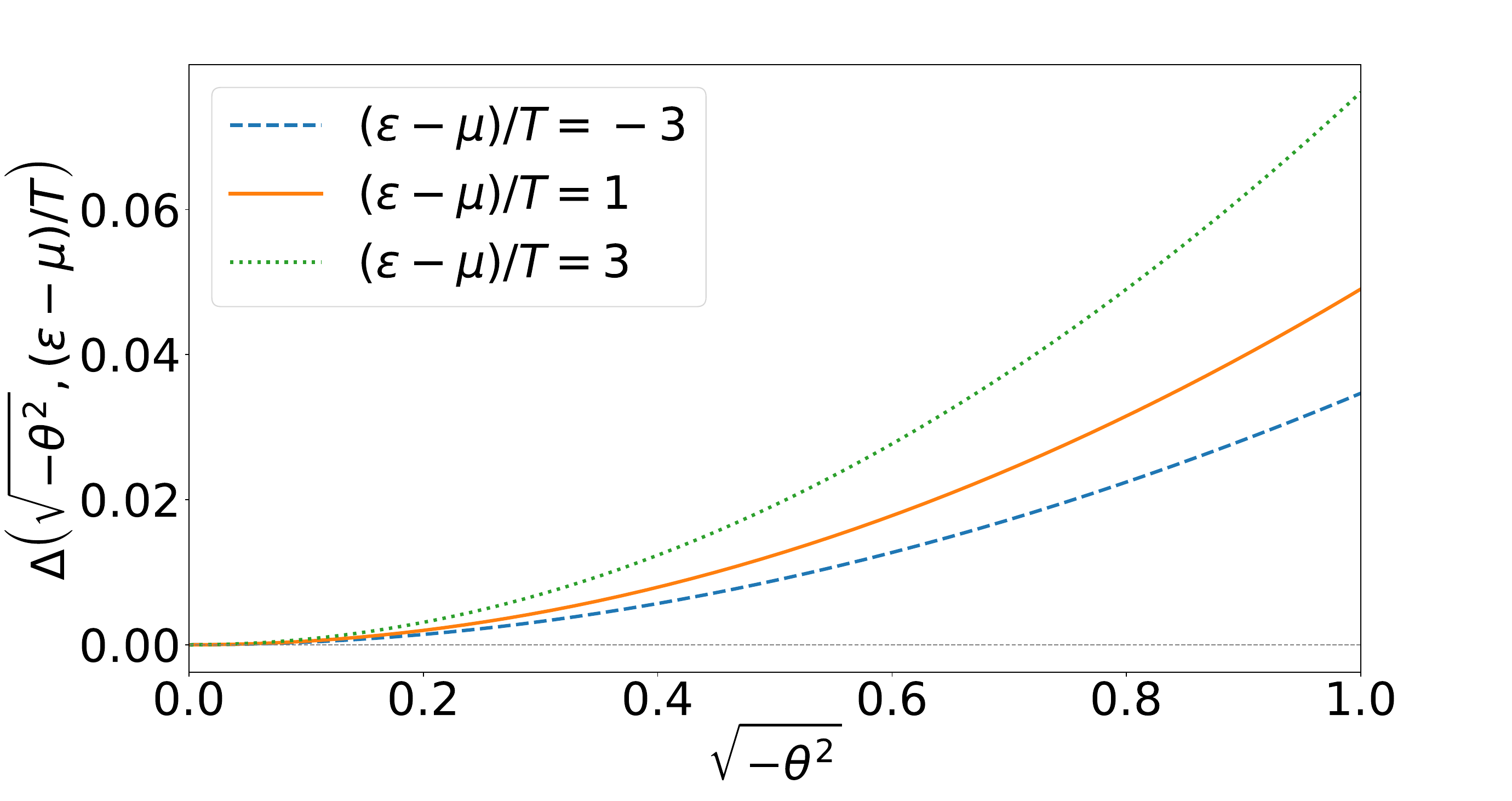}
    \includegraphics[width=0.49\textwidth]{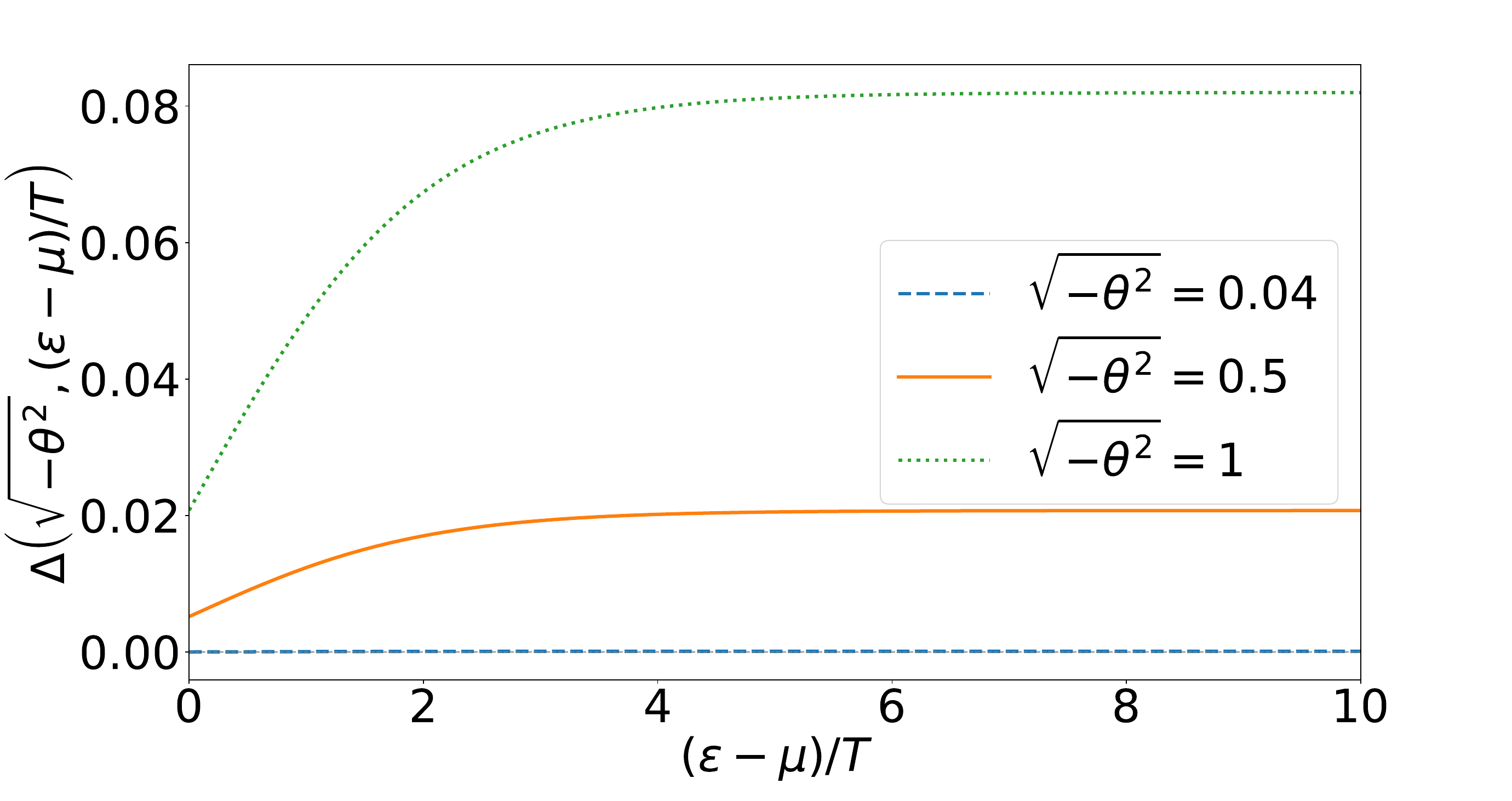}
    \caption{The behaviour of the relative difference between the exact polarization vector 
    and the linear approximation as a function of vorticity, temperature and energy and chemical 
    potential for $b\cdot p =\varepsilon/T$.}
    \label{fig:Delta}
\end{figure}

We can focus on the relative difference between the exact vorticity-induced spin vector eq. 
\eqref{eq:spin exact with s} 
and its linear approximation \eqref{eq: small vorticity massive} in global equilibrium. As the direction 
of the spin vector is given by $\theta^\mu$ in both formulae, the relative difference is the same for all 
components and reads:
\begin{equation*}
    \Delta\left(\sqrt{-\theta^2},b\cdot p-\zeta\right)=
    \frac{1}{P_{1/2}\left(\sqrt{-\theta^2},b\cdot p-\zeta\right)}
    \left(P_{1/2}\left(\sqrt{-\theta^2},b\cdot p-\zeta\right)-
    \frac{\sqrt{-\theta^2}}{2}\left(1-n_F(b\cdot p - \zeta)\right)\right).
\end{equation*}
In figure \ref{fig:Delta} we show $\Delta$ in terms of its arguments, having set $b\cdot p =\varepsilon/T$. It 
can be seen that, even for $\sqrt{-\theta^2}\sim 1$, the relative difference is less than $10\%$.

\begin{figure}
    \centering
    \includegraphics[width=0.49\textwidth]{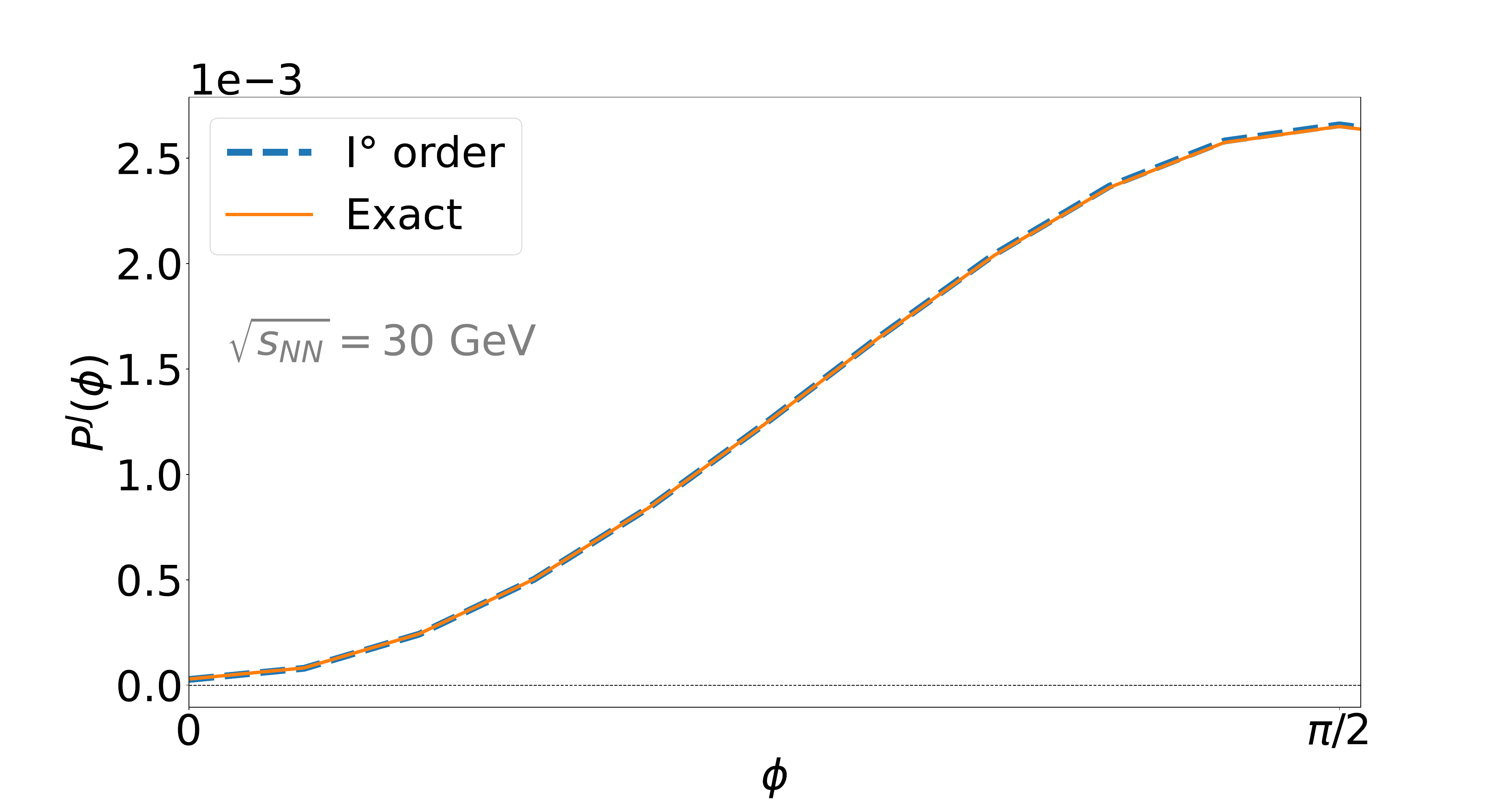}
    \includegraphics[width=0.49\textwidth]{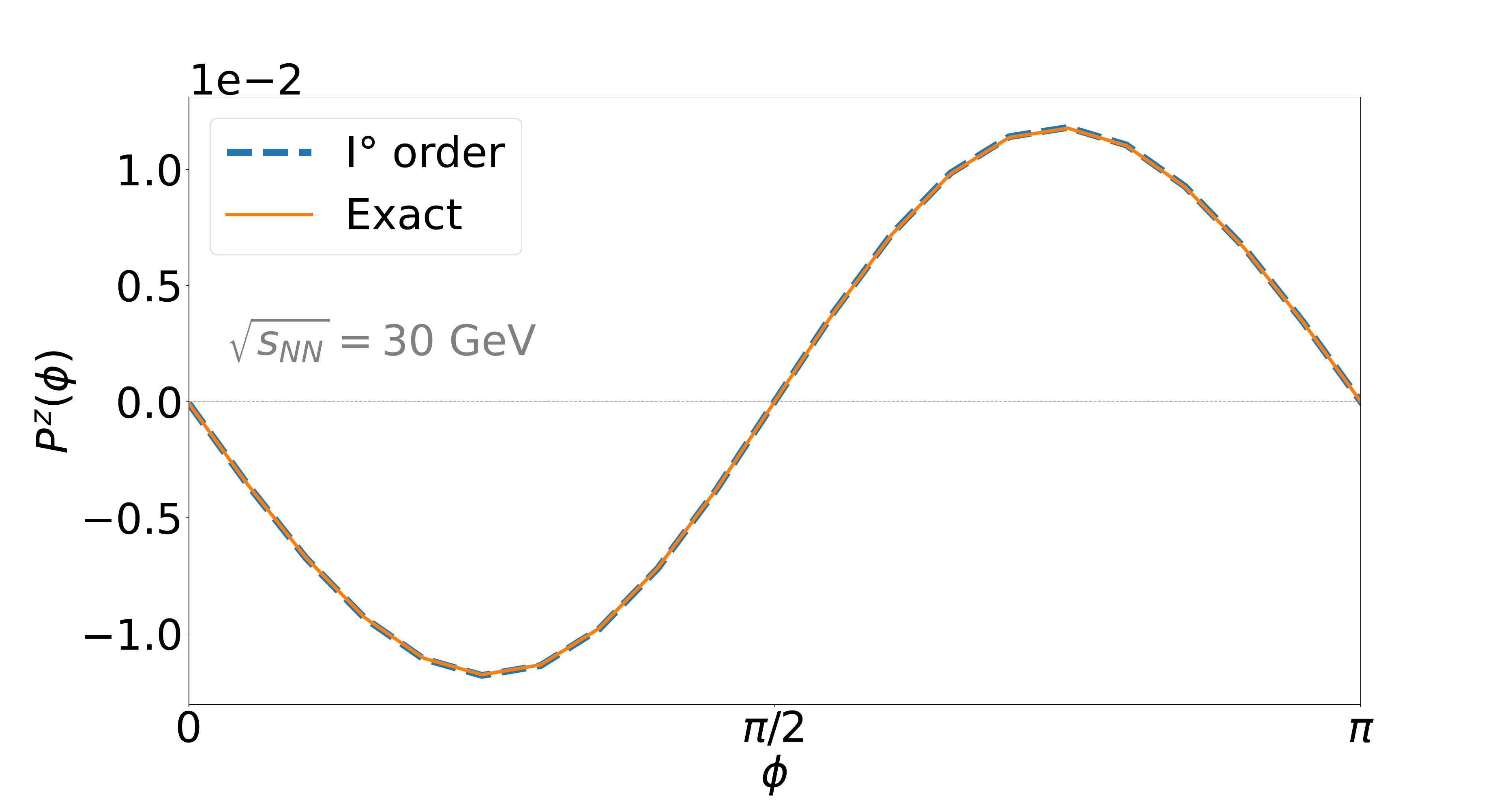}\\
    \includegraphics[width=0.49\textwidth]{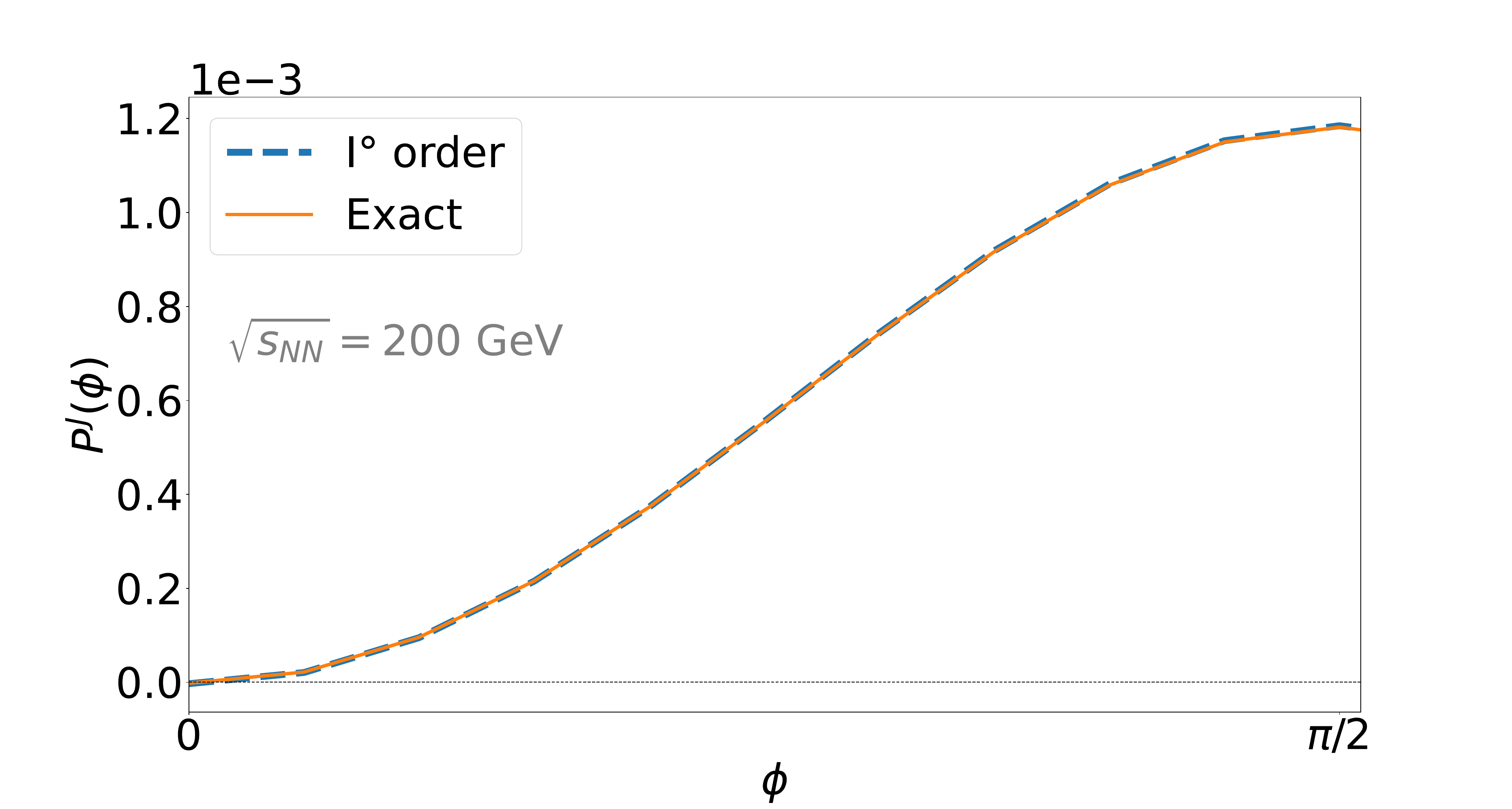}
    \includegraphics[width=0.49\textwidth]{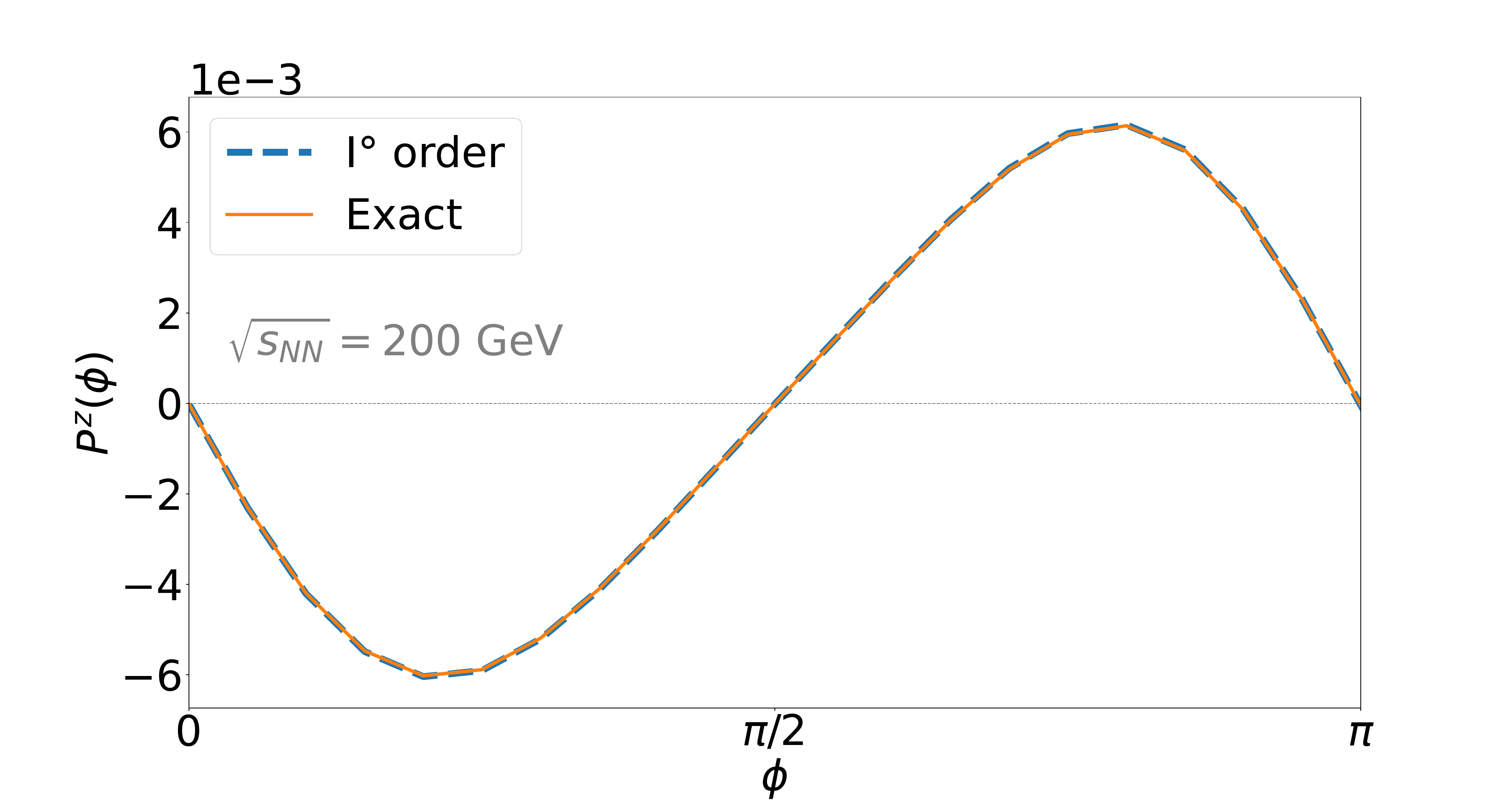}
    \caption{The components of the polarization vector along the angular momentum, $P^J$, and the beam 
    axis, $P^z$, are shown as functions of the azimuthal angle $\phi$ in the left and right panels respectively. 
    They are calculated at freeze-out in Au-Au collisions at $\sqrt{s_{NN}}=30$ (upper panels) 
    and $\sqrt{s_{NN}}=200$ GeV (lower panels). }
    \label{fig:local pol}
\end{figure}

At local equilibrium, thermal vorticity $\varpi$ can be promoted to a local variable, i.e. 
$\varpi=\varpi(x)$. Extending the formula \eqref{eq:linear spin vector vort} by using the 
equation \eqref{eq:spin exact with s}, the spin polarization vector of the $\Lambda$ 
induced by thermal vorticity at freeze-out turns out to be:
\begin{equation}\label{eq:pheno exact pol}
S^\mu(p)=-\frac{1}{4m}\epsilon^{\mu\nu\rho\sigma}p_\sigma\frac{\int\di\Sigma\cdot p\,
n_F\varpi_{\nu\rho}\, P_{1/2}\left(\sqrt{-\theta^2},
\frac{\varepsilon-\mu}{T}\right)/\sqrt{-\theta^2}}{\int\di\Sigma\cdot p\,n_F}.
\end{equation}
It should be emphasized that the above expression is not the exact spin vector at local equilibrium, 
as the contributions from thermal shear as well as from dissipative corrections and higher-order derivatives of the 
thermodynamic fields are not included. Nevertheless, the formula \eqref{eq:pheno exact pol} resums 
all the terms involving thermal vorticity and it is certainly a better approximation than 
\eqref{eq:linear spin vector vort} . 

We have evaluated the eq. \eqref{eq:pheno exact pol} by performing $3+1D$ hydrodynamic simulation of 
Au-Au collisions at $\sqrt{s_{NN}}=30$ and $\sqrt{s_{NN}}=200$ GeV with centrality $10-60\%$ by 
using the code vHLLE \cite{Karpenko:2013wva} for the hydrodynamic evolution and the integration over 
the freeze-out hypersurface. An averaged entropy density profile, generated by GLISSANDO v.2.702 
\cite{Rybczynski:2013yba}, is used as initial state. Only particles with zero rapidity have been taken 
into account.

Figure \ref{fig:local pol} shows the comparison between \eqref{eq:pheno exact pol} and the 
eq. \eqref{eq:linear spin vector vort}, the former being labelled as ``Exact'' and the latter as 
``I° order''. In particular, as it is customary, we study the azimuthal dependence $P^z$ and $P^J$, 
which are the projections of the polarization vector along the beam and the angular momentum directions 
respectively. We remind the reader that the polarization vector of Dirac fermions is twice the spin 
vector, $P^\mu=2S^\mu$.

Figure \ref{fig:local pol} confirms the expectations from fig. \ref{fig:Delta}, as the difference 
between the exact polarization and the linear approximation is tiny for the physical value of
thermal vorticity in relativistic heavy ion collisions at $30 < \sqrt{s_{\rm NN}}< 200$ GeV. 
It is possible that at lower energy, where thermal vorticity is larger \cite{Tsegelnik:2022eoz,Deng:2020ygd,Guo:2021udq,Jiang:2016woz,Ivanov:2020udj}, such corrections may play 
a more significant role.

\section{Summary and conclusions}

To summarize, we have derived the analytic formulae of the exact spin polarization vector and spin density 
matrix for massive and massless free fields at general global equilibrium with non-vanishing thermal vorticity. 
Our formulae are effectively a resummation of all higher-order corrections in thermal vorticity to the 
spin polarization vector and the spin density matrix. Furthermore, the unitary polarization bound 
is fulfilled.

We have developed the basic tools to study the polarization of massless particles, expressing the mean 
Pauli-Lubanski vector in terms of the spin density matrix and, for spin-1/2 particles, of the Wigner function. 
In agreement with the expectation, the mean Pauli-Lubanki vector is parallel to the four-momentum of 
the particle.   

We have studied the phenomenological implications of the improved formulae showing that the higher-order 
corrections to the spin polarization vector in thermal vorticity contribute marginally to the local 
polarization for $30$$<\sqrt{s_{NN}}<200$ GeV. For collisions with $\sqrt{s_{NN}}\sim 3-7$ GeV, where 
the vorticity is larger, they might be more significant. 

\section*{Acknowledgements}

A.P. acknowledges fruitful discussions with P. Aasha, V. Ambru\cb{s}, M. Buzzegoli, E. Grossi, D. Rischke 
and D. Wagner, as well as the kind hospitality of the Institute for Theoretical Physics, Goethe University, 
Frankfurt am Main (Germany), where part of this work was completed. A.P. also acknowledges the support of 
the fellowship ``Studio della polarizzazione nel plasma di QCD nelle collisioni nucleari di altissima energia'' 
by the University of Florence and the support of the Deutsche Forschungsgemeinschaft
(DFG, German Research Foundation) through the CRC-TR 211 `Strong-interaction matter under extreme conditions'– 
project number 315477589 – TRR 211. 

\bibliographystyle{ieeetr}
\bibliography{biblio}

\appendix
\section{Traces}\label{app: traces}

To compute the traces appearing in eqs. \eqref{eq: exact S massive} and \eqref{eq:exac PL massless}, we 
resort to the techniques used in ref. \cite{Prokhorov:2017atp}. First, we define the auxiliary variables: 
\begin{align}
z=\frac{\phi:\phi}{2}+i\frac{\phi:\tilde{\phi}}{2}, &&\bar{z}=\frac{\phi:\phi}{2}-i\frac{\phi:\tilde{\phi}}{2},
\end{align}
where $\widetilde{\phi}$ is the dual of $\phi$:
$$
\tilde{\phi}^{\mu\nu}=\frac{1}{2}\epsilon^{\mu\nu\rho\sigma}\phi_{\rho\sigma}.
$$
It is possible to show that the following identities hold \cite{Prokhorov:2017atp}:
\begin{align*}
 \tr(\Sigma^{\mu\nu}(\phi:\Sigma)^{2k+1})=&(\phi^{\mu\nu}+i\tilde{\phi}^{\mu\nu})z^k+(\phi^{\mu\nu}
 -i\tilde{\phi}^{\mu\nu})\bar{z}^k,\\
 \tr(\gamma_5\Sigma^{\mu\nu}(\phi:\Sigma)^{2k+1})=&(\phi^{\mu\nu}+i\tilde{\phi}^{\mu\nu})z^k-
 (\phi^{\mu\nu}-i\tilde{\phi}^{\mu\nu})\bar{z}^k,\\
 \tr(\Sigma^{\mu\nu}(\phi:\Sigma)^{2k})=&0,\\ \tr(\gamma_5\Sigma^{\mu\nu}(\phi:\Sigma)^{2k})=&0.
\end{align*}
The above equations allow us to determine the traces involved in the calculations of sec. \ref{sec: exact pol fermions}. 
For example one has:
\begin{align*}
     \tr(\exp [-in\phi:\Sigma/2]))
     &=\sum_{k=0}^\infty\frac{n^k(-i)^k}{2^{k}k!}\phi_{\mu\nu}\tr(\Sigma^{\mu\nu}(\phi:\Sigma)^{k-1})\\ 
     &=\sum_{k=-1}^\infty\frac{n^{2k+2}(-i)^{2k+2}}{2^{2k+2}(2k+2)!}\phi_{\mu\nu}\tr(\Sigma^{\mu\nu}(\phi:\Sigma)^{2k+1})\\
     &=\sum_{k=-1}^\infty\frac{n^{2k+2}(-i)^{2k+2}}{2^{2k+1}(2k+2)!}(z^{k+1}+\bar{z}^{k+1})\\
     &=2\cos\left(\frac{n\sqrt{z}}{2}\right)+2\cos\left(\frac{n\sqrt{\bar{z}}}{2}\right).
 \end{align*}
 Similarly:
 \begin{align*}
 \tr(\gamma_5\exp [-in\phi:\Sigma/2]))&=2\cos\left(\frac{n\sqrt{z}}{2}\right)-2\cos\left(\frac{n\sqrt{\bar{z}}}{2}\right),\\
 \tr(\Sigma^{\mu\nu}\exp [-in\phi:\Sigma/2])&=-i(\phi^{\mu\nu}+i\tilde{\phi}^{\mu\nu})\frac{\sin\left(\sqrt{z}/2\right)}
 {\sqrt{z}}-i(\phi^{\mu\nu}-i\tilde{\phi}^{\mu\nu})\frac{\sin\left(\sqrt{\bar{z}}/2\right)}{\sqrt{\bar{z}}},\\
\tr(\gamma_5\Sigma^{\mu\nu}\exp [-in\phi:\Sigma/2])&=-i(\phi^{\mu\nu}+i\tilde{\phi}^{\mu\nu})
\frac{\sin\left(\sqrt{z}/2\right)}{\sqrt{z}}+i(\phi^{\mu\nu}-i\tilde{\phi}^{\mu\nu})
\frac{\sin\left(\sqrt{\bar{z}}/2\right)}{\sqrt{\bar{z}}}.
\end{align*}
These formulae hold for any $\phi$, and in the general $z$ and $\bar{z}$ are complex numbers. However, 
the cases of interest are always such that $\phi:\widetilde{\phi}=0$ and $z=\bar{z}=\phi:\phi/2$, see 
for instance eqs. \eqref{eq: phi:phi} and \eqref{eq: phi:phi massless}. 

For $z=\bar{z}$ the above traces simplify, and one has:
\begin{subequations}
\begin{align*}
\tr(\exp [-in\phi:\Sigma/2]))
     &=4\cos\left(\frac{n\sqrt{z}}{2}\right),\\
 \tr(\gamma_5\exp [-in\phi:\Sigma/2]))&=0,\\
 \tr(\Sigma^{\mu\nu}\exp [-in\phi:\Sigma/2])&=-2i\phi^{\mu\nu}\frac{\sin\left(\sqrt{z}/2\right)}{\sqrt{z}},\\
\tr(\gamma_5\Sigma^{\mu\nu}\exp [-in\phi:\Sigma/2])&=2\tilde{\phi}^{\mu\nu}\frac{\sin\left(\sqrt{z}/2\right)}{\sqrt{z}}.
\end{align*} 
\end{subequations}
Writing the product of two gamma matrices in terms of their commutator and anticommutator, 
$[\gamma^\mu,\gamma^\nu]=-4i\Sigma^{\mu\nu}$ and $\{\gamma^\mu,\gamma^\nu\}=2g^{\mu\nu}$ respectively, 
we have:
\begin{align*}
    \tr(\gamma^\nu \gamma^\mu \exp [-in\phi:\Sigma/2]) &= g^{\mu\nu}\tr(\exp [-in\phi:\Sigma/2])
    +2i\tr(\Sigma^{\mu\nu}\exp [-in\phi:\Sigma/2]),\\
    \tr(\gamma^\nu \gamma^\mu \gamma_5 \exp [-in\phi:\Sigma/2]) 
    &=g^{\mu\nu}\tr(\gamma_5\exp [-in\phi:\Sigma/2])+2i\tr(\Sigma^{\mu\nu}\gamma_5\exp [-in\phi:\Sigma/2]),
\end{align*}
and finally we find:
\begin{subequations}\label{A and A5 constraint}
\begin{align}
\tr(\exp [-in\phi:\Sigma/2])) =&4\cos\left(\frac{n\sqrt{z}}{2}\right),\\
 \tr(\gamma^\nu \gamma^\mu \exp [-in\phi:\Sigma/2])=&4g^{\mu\nu}\cos\left(\frac{n\sqrt{z}}{2}\right)
 +4\phi^{\mu\nu}\frac{\sin\left(\sqrt{z}/2\right)}{\sqrt{z}},\\
\tr(\gamma^\nu \gamma^\mu \gamma_5 \exp [-in\phi:\Sigma/2])=
&4i\tilde{\phi}^{\mu\nu}\frac{\sin\left(\sqrt{z}/2\right)}{\sqrt{z}},
\end{align} 
\end{subequations}
which are precisely the identities \eqref{eq: identities traces}.

\section{Little group for massless particles}\label{app: constraint}

In this section, the form of the tensor $\phi$ fulfilling $\phi^{\mu\nu} p_\nu = 0$ for a light-like 
momentum $p$ is obtained. 
To begin with, we find the most general decomposition of an anti-symmetric tensor using the
the basis $\{p,q,n_1,n_2\}$, with $p^2=q^2=0$, which has been introduced in sec.
~\ref{sec: Polarization QFT}. It can be shown that two vectors $h$ and $y$ exist, with
$h \cdot q =0$ and $y\cdot p=0$, such that:
\begin{equation*}
    \phi^{\mu\nu}=\epsilon^{\mu\nu\rho\sigma}\frac{h_\rho p_\sigma}{p\cdot q}+y^\mu q^\nu-y^\nu q^\mu.
\end{equation*}
Their existence can be proved by inverting the above relation. Indeed, contracting $\phi^{\mu\nu}$
with $p_\nu$ and taking into account that $y\cdot p=0$:
\begin{equation*}
    \phi^{\mu\nu}p_\nu = y^\mu p\cdot q \Rightarrow y^\mu = \frac{\phi^{\mu\nu}p_\nu}{p\cdot q},
\end{equation*}
what is consistent with the requirement $y\cdot p=0$. Furthermore, if $h\cdot q=0$:
\begin{equation*}
    \epsilon_{\mu\nu\alpha\beta}\phi^{\mu\nu}q^\alpha=
    \epsilon_{\mu\nu\alpha\beta}\epsilon^{\mu\nu\rho\sigma}\frac{h_\rho p_\sigma}{p\cdot q} q^\alpha 
    = -2(\delta^{\rho}_{\alpha}\delta^{\sigma}_\beta-\delta^{\rho}_{\beta}\delta^{\sigma}_\alpha)
    h_\rho \frac{p_\sigma q^\alpha}{p\cdot q} = 2h_\beta ,
\end{equation*}
which leads to:
\begin{equation*}
h^\mu=-\frac{1}{2}\epsilon^{\mu\nu\rho\sigma}\phi_{\nu\rho}q_\sigma,    
\end{equation*}
which is again consistent with the requirement $h\cdot q=0$.

Using the above decompostion, the solution of $\phi^{\mu\nu}p_\nu=0$ simply yields $y=0$, 
which implies that the little group of massless particles is generated by tensors $\phi$ 
parametrized as:
\begin{equation*}
    \phi^{\mu\nu}=\epsilon^{\mu\nu\rho\sigma}\frac{h_\rho p_\sigma}{p\cdot q},
\end{equation*}
with $h\cdot q=0$, that is the equation \eqref{eq:constrained vort massless}.

\section{The little-group transformations with $\Lambda p = p$}\label{app:wigner phase}

Our goal is to calculate the Wigner rotation, for massive and massless particles, for Lorentz 
transformations such that $\Lambda p = p$. For this set of transformations, the Wigner rotation 
can be written as follows:
\begin{equation}\label{eq: simplified wigner rot}
  W(\Lambda,p)=[\Lambda p]^{-1}\Lambda[p]=[p]^{-1}\Lambda[p]=
 \exp\left[-\ii\frac{\phi_{\mu\nu}}{2}[p]^{-1}J^{\mu\nu}[p]\right]=
 \exp\left[-\ii\frac{{\phi_0}_{\mu\nu}}{2}J^{\mu\nu}\right],
\end{equation}
where we have used the condition $\Lambda p = p$, the transformation rules of the generators $J^{\mu\nu}$, 
and denoted 
$$
{\phi_0}_{\mu\nu}=\phi_{\rho\sigma}[p]^{\rho}_{\ \mu}[p]^{\sigma}_{\ \nu} \;\; .
$$ 
For massive particles, using the parametrization \eqref{eq: phi:phi} and the properties of the 
Levi-Civita symbol, one has:
\begin{equation}\label{eq: conto phi0}
    {\phi_0}_{\mu\nu}=\phi_{\rho\sigma}[p]^{\rho}_{\ \mu}[p]^{\sigma}_{\ \nu}=\frac{1}{m}\underbrace{\epsilon_{\rho\sigma\lambda\tau}[p]^{\rho}_{\ \mu}[p]^{\sigma}_{\ \nu}[p]^{\lambda}_{\ \lambda'}[p]^{\tau}_{\ \tau'}}_{=\epsilon_{\mu\nu\lambda'\tau'}}{[p]^{-1}}^{\lambda'}_{\ \alpha}{[p]^{-1}}^{\tau'}_{\ \gamma}\xi^\alpha p^\gamma=\frac{1}{m}\epsilon_{\mu\nu\rho\sigma}{\xi_0}^\rho \mathfrak{p}^\sigma,
\end{equation}
where $\xi_0^\mu = {[p]^{-1}}^{\mu}_{\ \nu}\xi^\nu$, the vector $\xi$ being defined by \eqref{eq:constraint global pol}, and $\mathfrak{p}^\mu={[p]^{-1}}^{\mu}_{\ \nu}p^\nu$. 
Plugging this expression in \eqref{eq: simplified wigner rot} and going to the representation
$D^S$ of the rotation group, one obtains:
\begin{equation*}
    D^S(W(\Lambda,p))=\exp\left[-\ii\frac{{\phi_0}_{\mu\nu}}{2}D^S(J^{\mu\nu})\right]
    =\exp[-i\bm{\xi_0}\cdot D^S(\textbf{J})],
\end{equation*}
which gives equation \eqref{eq: wigner rot massiva simplified}.

We now move to the massless case. Similarly as in \eqref{eq: conto phi0} and using the parametrization \eqref{eq:constrained vort massless}, the tensor $\phi_0$ reads:
\begin{equation*}
    \phi^{\mu\nu}_0=\frac{1}{\mathfrak{p}\cdot \mathfrak{q}}\epsilon^{\mu\nu\rho\sigma}\mathfrak{h}_\rho \mathfrak{q}_\sigma,
\end{equation*}
where $\mathfrak{h}^\mu={[p]^{-1}}^{\mu}_{\ \nu}h^\nu$, the vector $h$ being defined in \eqref{eq:constrained vort massless}.
Using \eqref{eq: simplified wigner rot} and the decomposition \eqref{eq:constrained vort massless}:
\begin{equation*}
    W(\Lambda,p) = \exp\left[-\ii\frac{\phi_0:J}{2} \right]=
    \exp\left[-\ii \frac{\mathfrak{h}\cdot \Pi(\mathfrak{p})}{\mathfrak{q}\cdot\mathfrak{p}}\right]
    =\exp\left[\ii\frac{\mathfrak{h}\cdot\mathfrak{p}}{\mathfrak{q}\cdot\mathfrak{p}} h(\mathfrak{p}) + \ii\frac{\mathfrak{h}\cdot\mathfrak{n}_1}{\mathfrak{q}\cdot\mathfrak{p}}\Pi_1(\mathfrak{p})
    +\ii\frac{\mathfrak{h}\cdot\mathfrak{n}_2}{\mathfrak{q}\cdot\mathfrak{p}}\Pi_2(\mathfrak{p})\right].
\end{equation*}
The generator $h(\mathfrak{p})$ coincides with ${\rm J}^3$ (see section~\ref{sec: Polarization QFT}).
Since, according to the algebra \eqref{eq:algebra h massless}, the commutator between $h$ and 
$\Pi_{1,2}$ can be written in terms of $\Pi_{1,2}$, it can be realized that the Baker-Cambpell-Haussdorf 
formula for the factorization of exponentials of operators implies the existence of $a_1$, $a_2$ 
such that:
\begin{equation*}
  \exp\left[\ii\frac{\mathfrak{h}\cdot\mathfrak{p}}{\mathfrak{q}\cdot\mathfrak{p}} h(\mathfrak{p}) + \ii\frac{\mathfrak{h}\cdot\mathfrak{n}_1}{\mathfrak{q}\cdot\mathfrak{p}}\Pi_1(\mathfrak{p})
    +\ii\frac{\mathfrak{h}\cdot\mathfrak{n}_2}{\mathfrak{q}\cdot\mathfrak{p}}\Pi_2(\mathfrak{p})\right]
    =\exp\left[\ii a_1 \Pi_1(\mathfrak{p})+\ii a_2 \Pi_2(\mathfrak{p})\right]\exp\left[\ii \frac{\mathfrak{h}\cdot\mathfrak{p}}{\mathfrak{q}\cdot\mathfrak{p}} h(\mathfrak{p})\right],
\end{equation*}
The representation of such transformation onto the Hilbert space of on-shell states 
(see section~\ref{sec: Polarization QFT}) is such that the first exponential is the identity due to \eqref{eq s1|p>=0}, and
only the rightmost exponential contributes. We thus have eq. \eqref{wigrotmassless} 
with:
$$
\vartheta(\Lambda,p)=\frac{\mathfrak{h}\cdot \mathfrak{p}}{\mathfrak{q}\cdot\mathfrak{p}}=
\frac{h\cdot p}{q\cdot p}=\eta
$$
where we used the definition \eqref{def: eta massless}.

\end{document}